\documentclass[lettersize,journal]{IEEEtran}
\usepackage{amsmath,amsfonts}
\usepackage{algorithmic}
\usepackage{array}
\usepackage[caption=false,font=normalsize,labelfont=sf,textfont=sf]{subfig}
\usepackage{textcomp}
\usepackage{stfloats}
\usepackage{url}
\usepackage{verbatim}
\usepackage{hyperref}
\usepackage{graphicx}
\usepackage{ulem}
\usepackage{floatrow}
\usepackage{xcolor}

\usepackage{printlen}

\usepackage{algorithm}
\usepackage{bbm}

\newcommand{\dav}[1]{\textcolor{black}{#1}}

\usepackage{multirow}

\usepackage{caption}   
\usepackage{subfig}    
\usepackage[style=ieee]{biblatex}
\def\BibTeX{{\rm B\kern-.05em{\sc i\kern-.025em b}\kern-.08em
    T\kern-.1667em\lower.7ex\hbox{E}\kern-.125emX}}
\usepackage{balance}

\usepackage{soul}

\newcommand{\ve}[1]{ {\mathbf{#1}} }
\def\bal#1{\begin{align}#1\end{align}}

\usepackage{booktabs}

\begin{document}
\sloppy
\title{Enhancing time-frequency resolution with optimal transport and barycentric fusion of multiple spectrograms}

\author{David Valdivia, Elsa Cazelles, Cédric Févotte, \IEEEmembership{Fellow,~IEEE,}
\thanks{The authors are with IRIT, University of Toulouse, CNRS, France (email: firstname.lastname@irit.fr). This work is supported by the AI Interdisciplinary Institute ANITI, funded by the France 2030 program under the grant agreement ANR-23-IACL-0002, and by the France 2030 program PEPR, under the grant agreement ANR-23-PEIA-0004. The generative AI {\em Le Chat} by Mistral (used under CNRS license) was used to improve grammar, spelling, tone, or formatting of human-written text throughout the manuscript.} 
}
\maketitle

\begin{abstract}
Time-frequency representations, such as the short-time Fourier transform (STFT), are fundamental tools for analyzing non-stationary signals. However, their ability to achieve sharp localization in both time and frequency is inherently limited by the Gabor-Heisenberg uncertainty principle. In this paper, we address this limitation by introducing a method to generate super-resolution spectrograms through the fusion of two or more spectrograms with varying resolutions. Specifically, we compute the super-resolution spectrogram as the barycenter of input spectrograms using optimal transport (OT) divergences. Our method does not require the input spectrograms to share the same time-frequency grid. The input spectrograms can be computed using any STFT parameters, and the resulting super-resolution spectrogram can be defined on an arbitrary user-specified grid. We explore various OT divergences based on different transportation costs.
 Notably, we introduce a novel transportation cost that preserves time-frequency geometry while significantly reducing computational complexity compared to standard Wasserstein barycenters. We adopt the unbalanced OT framework and derive a new block majorization-minimization algorithm for efficient barycenter computation. 
 We validate the proposed method on controlled synthetic signals and recorded speech using both quantitative and qualitative evaluations. The results show that our approach combines the best localization properties of the input spectrograms \dav{while maintaining a smooth energy distribution, and preserving computational efficiency.}
\end{abstract}

\begin{IEEEkeywords}
Time-frequency analysis, optimal transport,  super-resolution, audio signal processing, fusion of spectrograms
\end{IEEEkeywords}


\section{Introduction} \label{section:introduction}

\begin{figure}[ht!]
\centering
\subfloat[]{%
  \includegraphics[width=\columnwidth]{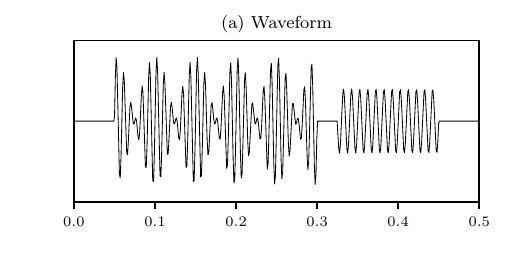}%
}\\[-12mm]
\subfloat[]{%
  \includegraphics[width=\columnwidth]{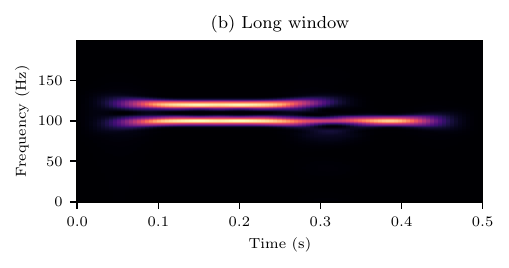}%
}\\[-12mm]
\subfloat[]{%
  \includegraphics[width=\columnwidth]{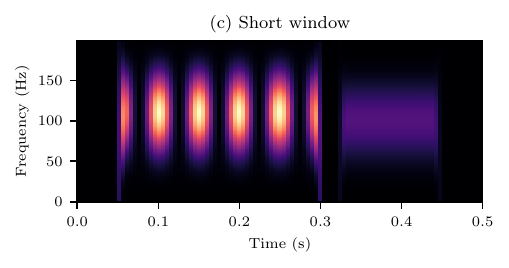}%
}\\[-12mm]
\subfloat[]{%
  \includegraphics[width=\columnwidth]{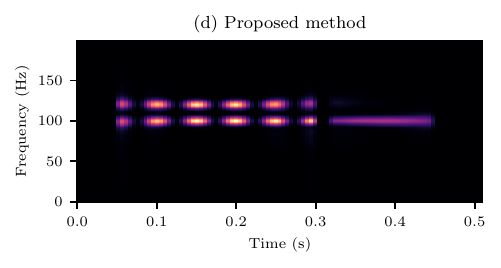}%
  \label{fig:three}%
}\\[-10mm]
\caption{
 Impact of window length on spectrogram resolution. (a) Time-domain waveform: Two sinusoidal components at $100\,\mathrm{Hz}$ and $120\,\mathrm{Hz}$ interfere, creating a beating pattern. After a brief pause, only the $100\,\mathrm{Hz}$ component persists. (b) Spectrogram computed with a long window \dav{(\(180\,\mathrm{ms}\))}, effectively separating the sinusoidal components but blurring the temporal structure, including the pause and beating pattern. (c) Spectrogram computed with a short window \dav{(\(20\,\mathrm{ms}\))}, preserving the temporal localizations  but failing to resolve the individual sinusoidal frequencies. (d) Super-resolution spectrogram obtained with our method, preserving both the temporal and frequency localizations. 
 }
\label{fig:spectrogram-resolution}
\end{figure}

\IEEEPARstart{N}ON-STATIONARY signals, such as audio recordings and EEG data, are ubiquitous in real-world applications of signal processing. Due to their time-varying spectral properties, traditional frequency-only analysis tools like the Fourier transform (FT) fail to fully capture their dynamic characteristics. Time-frequency representations (TFRs) thus offer a more effective analytical framework. Among TFRs, the short-time Fourier transform (STFT) and its power magnitude representation, the spectrogram, are widely adopted due to their simplicity and computational efficiency \cite[Chapter~10.3]{oppenheim1999discrete}. The STFT partitions a signal into a series of windowed segments, each analyzed using the FT. However, this approach is fundamentally constrained by the Gabor-Heisenberg uncertainty principle \cite{mallat1999wavelet}, which prevents simultaneous sharp localization in both time and frequency. The choice of window duration introduces a critical trade-off: a long window improves frequency resolution but blurs temporal events, while a short window captures transient phenomena at the expense of frequency precision. This limitation is illustrated in Fig.~\ref{fig:spectrogram-resolution}. To overcome this challenge, in this paper we propose a novel method to enhance time-frequency (t-f) resolution by fusing multiple spectrograms with varying resolutions, using optimal transport (OT).

\subsection{Related work in time-frequency enhancement}

\dav{Various methods have been developed to obtain super-resolution TFRs meant to achieve simultaneously good temporal and frequency localization, beyond the Gabor–Heisenberg uncertainty limit.} One of the earliest approaches is proposed in \cite{cheung1991combined}, where the authors introduce a fusion method based on the pointwise geometric mean of two spectrograms computed with different window lengths. For direct combination, the t-f grids of the two spectrograms must align, which is achieved by setting a common hop size and frequency spacing. This results in spectrograms with denser time or frequency sampling than typical default settings (e.g., a hop size of half the window length and minimal frequency sampling). However, this increased redundancy amplifies smearing and boundary artifacts.

A model-based fusion method is proposed in \cite{nam2010super,Leplat2022} using Nonnegative Matrix Factorization (NMF). Here, the two input spectrograms are modeled as the product of two non-negative matrices: one representing a dictionary of frequency patterns and the other an activation matrix. The spectrograms are assumed to share common latent factors that are degraded in either time or frequency resolution by user-defined operators. An optimization algorithm recovers the high-resolution latent factors by solving a joint NMF problem, and the super-resolution spectrogram is obtained as the product of these factors. While this approach yields satisfactory results for certain signals (e.g., musical signals), its performance depends heavily on the validity of the NMF low-rank assumption, making it less suitable for complex signals like speech or environmental sounds. A related method is presented in \cite{kirbiz2014multiresolution}, where tensor factorization is employed to fuse spectrograms with varying resolutions.

Besides fusion, t-f analysis includes handcrafted representations and post-processing techniques designed to improve resolution for specific signal classes. For example, the Wigner-Ville distribution provides precise t-f localization for signals like chirps \cite{boashash2015time,Flandrin1998}. However, its utility is limited by cross-terms that arise in multi-component signals, reducing interpretability, though these cross-terms can be mitigated using Cohen’s class of distributions \cite{Cohen1989,shafi2009techniques} for an in-depth review of such methods. Another post-processing technique, t-f reassignment, sharpens TFRs by relocating energy to estimated t-f centroids \cite{flandrin2018time}. Finally, multiscale methods, such as the continuous wavelet transform \cite{grossmann1990reading} and its extensions like the superlet transform \cite{moca2021time}, combine multiple analysis windows to improve resolution across scales.

\subsection{Our approach}

We propose a super-resolution approach that combines spectrograms of different t-f resolution using OT barycenters. OT is a mathematical framework for comparing probability distributions under geometric constraints \cite{peyre2020computationalot}. The geometry of the problem is encoded through cost matrices, which define the cost of transporting probability mass across a space. OT barycenters generate intermediary distributions that effectively average the characteristics of the input distributions.

In our work, spectrograms are treated as energy distributions supported on a set of t-f points. Given two spectrograms of the same signal computed with different window lengths, and thus different t-f resolutions, our approach involves computing an OT barycenter of the two input t-f distributions. The resulting barycentric spectrogram retains the better frequency resolution provided by the spectrogram with the longest window and the better temporal resolution provided by the other, yielding a super-resolution spectrogram. 
In contrast with some other methods, we eliminate the need to align spectrograms on a common grid by leveraging the t-f geometry. This avoids unnecessary refinement of the t-f discretization (e.g., additional time frames or frequency bins). Furthermore, the target support of the barycentric spectrogram can be chosen arbitrarily. In this paper, we focus on a simple construction in which the target support is a grid built from the frequency sampling of the high frequency resolution spectrogram and the temporal sampling of the high temporal resolution spectrogram. However, more general settings can be considered, and we provide such an example in the Supplementary Material.

The OT barycenter depends on the choice of a transportation cost between t-f points. We consider a traditional cost, yielding Wasserstein barycenters, and introduce a novel cost function that complies with the t-f geometry. As detailed in Section \ref{section:spec-fusion}, this leads us to adopt the unbalanced optimal transport (UOT) framework, and we design a novel block-descent majorization-minimization (MM) algorithm for computing UOT barycenters.
In summary, our work makes the following contributions:

\begin{itemize}
\item We formulate super-resolution via spectrogram fusion and fixed-support OT barycenter computation, accommodating spectrograms defined on different t-f grids.
\item We design structured transportation cost matrices that yield both improved results and significantly reduced runtimes compared to traditional Wasserstein barycenters.
\item We propose an efficient algorithm to compute the UOT barycenter of distributions with different supports, using block-descent MM (with potential impact that is beyond the scope of this paper).
\item We evaluate the proposed method on controlled synthetic signals and speech data, through quantitative and qualitative assessment of t-f localization \dav{and energy distribution}.
\end{itemize}

While OT is widely used in image processing and machine learning, its application in signal processing remains somewhat limited, and our work advances this direction. In our previous work \cite{valdivia2025audio}, we considered audio interpolation between two different input sounds by computing an OT barycenter between their spectrograms (supported by a common t-f grid), followed by time-domain signal reconstruction using the Griffin–Lim algorithm \cite{griffin1984signal}. Note that this setting differs from the one considered here, where we consider OT barycenters of spectrograms of the same sound with different t-f supports, and aim for super-resolution t-f analysis. Our previous work \cite{valdivia2025audio} built on the OT-based portamento method of \cite{Henderson2019}. In \cite{Flamary2016,elvander2017using}, OT is used for pitch estimation in music signals by matching the observed spectrum to an ideal harmonic distribution. Other applications of OT in signal processing include model-based spectral analysis \cite{Elvander2023}, audio similarity quantification \cite{cazelles2020wasserstein,Fabiani2024}, room transfer estimation \cite{Bjoerkman2025}, and blind source separation \cite{fabiani2025joint}.

The paper is organized as follows. Section \ref{section:preliminaries} introduces notations and background on OT and UOT. Section \ref{section:spec-fusion} presents our OT-based framework for spectrogram fusion. Section \ref{section:algorithm} describes our new algorithm for computing UOT barycenters. Sections \ref{sec:synth-signals} and \ref{sec:real-signals} report experimental results on synthetic and speech signals, respectively. Though we will mostly focus on the fusion of two spectrograms for simplicity, our method can straightforwardly accommodate the fusion of more spectrograms, as discussed in Section~\ref{sec:multi}. Furthermore, we provide in the Supplementary Material an example of target support customization where the target frequency grid is set to the non-uniform mel frequency scale \cite{umesh1999fitting}. The code used to reproduce the figures and experiments is available online.\footnote{ \url{https://github.com/davidvaldiviad/fusion-ot}} The repository also includes a tutorial for computing super-resolution spectrograms.


\section{Preliminaries} \label{section:preliminaries}

\subsection{Notations}

Let $\mathbb{R}_{+}$ represent the set of non-negative real numbers, $\mathbb{R}_{++}$ the set of positive real numbers, and $\mathbb{N}$ the set of integers. Matrices are written in bold uppercase, e.g., $\mathbf{X} \in \mathbb{R}^{M \times N}$, and vectors in bold lowercase, e.g., $\mathbf{a} \in \mathbb{R}^{I}$. Their entries are written in regular font, i.e., $X_{mn}$ and $a_i$. For two matrices (or vectors) $\mathbf{A} \in \mathbb{R}^{M \times N}$ and $\mathbf{B} \in \mathbb{R}^{M \times N}$, their entrywise multiplication is denoted as $\mathbf{A} \odot \mathbf{B}$ and entrywise division as $\frac{\mathbf{A}}{\mathbf{B}}$. For a given vector $\mathbf{a} \in \mathbb{R}^I$, $\mathrm{diag}(\mathbf{a})$ represents the diagonal matrix with diagonal entries $a_i$, $i=1,\dots,I$. For $I \in \mathbb{N}$, $\mathbf{1}_I$ denotes the column vector of size $I$ with all entries equal to one. Similarly, $\mathbf{1}_{M\times N}$ denotes the $M \times N$ matrix with all entries equal to one. Given a matrix $\mathbf{X} \in \mathbb{R}^{M \times N}$, its column-wise vectorization is defined as
\bal{
\mathrm{vec}(\mathbf{X}) = [X_{11} \dots X_{M1} \dots X_{MN}]^\top \in \mathbb{R}^{MN}.
\label{eq:vectorization}
}
{This vectorization underlies a mapping between $\mathbb{R}^{M \times N}$ and $\mathbb{R}^{MN}$ that we denote by}
\begin{equation}
    \pi(m, n) = i.
\label{eq:mapping}
\end{equation}
The mapping relates the entries $X_{mn}$ and $\mathrm{vec}(\mathbf{X})_i$. 

For a set of points $\mathcal{S} = \{s_i\}_{i=1}^{I}\subseteq \mathbb{R}^D$, $\mathcal{M}_{+}(\mathcal{S})$ denotes the set of non-negative discrete distributions (or measures) on $\mathcal{S}$. By definition, $\alpha \in \mathcal{M}_{+}(\mathcal{S})$ can be expressed as
\bal{
\alpha = \sum_{i = 1}^I a_{i}\delta_{s_{i}},
}
where the vector $\mathbf{a} \in \mathbb{R}_+^{I}$ denotes the {\em weights} of $\alpha$ and $\delta_{s_i}$ denotes the Dirac measure at location $s_i$. The set $\mathcal{S}$ is called the {\em support} of $\alpha$. We denote by $\mathcal{M}_{+}^{1}(\mathcal{S})$ the set of discrete {\em probability} distributions on $\mathcal{S}$, i.e., the set of distributions with weights $\mathbf{a} \in \mathbb{S}^I$, where $\mathbb{S}^I$ is the probability simplex defined as
\bal{
\mathbb{S}^I = \left\{ \mathbf{a} \in \mathbb{R}_{+}^I \;|\; \sum_{i = 1}^I a_{i} = 1 \right\}.
}
In optimal transport, the weights are also commonly referred to as {\em mass}. In this paper, we will also sometimes refer to them as {\em energy}, a term that arises naturally in our signal processing context.

Finally, for a non-negative vector $\mathbf{a} \in \mathbb{R}_{+}^I$ and a positive vector $\mathbf{b} \in \mathbb{R}_{++}^I$, the generalized Kullback-Leibler (KL) divergence is defined as
\bal{
\mathrm{KL}(\mathbf{a}, \mathbf{b}) = \sum_{i = 1}^{I} a_{i} \log\left(\frac{a_{i}}{b_{i}}\right) - a_{i} + b_{i},
\label{eq:kl}}
with the convention $0 \cdot \log(0) = 0$. The KL divergence is a natural measure of fit between non-negative vectors, particularly between probability vectors.

\subsection{The spectrogram}

Let $\mathbf{y}\in\mathbb{R}^L$ be a temporal signal sampled at frequency $f_s$ (in Hertz). To analyze its time-varying spectral properties, we consider the short-time Fourier transform (STFT). Given a window $\mathbf{g} \in \mathbb{R}^W$ and a hop size $H \in \mathbb{N}$, the STFT first partitions the signal into successive windowed segments of size $W$ (referred to as {\em frames} in the following), shifted by $H$ samples. Then a $N_{\mathrm{fft}}$-point discrete Fourier transform (DFT) is applied to every frame, with $N_{\mathrm{fft}}\geq W$. For simplicity and without loss of generality, we consider that $W$ and $N_{\mathrm{fft}}$ are both even-valued. As such, the STFT $\mathbf{Y} \in \mathbb{C}^{M \times N}$ of $\ve{y}$ is defined entrywise as 
\begin{equation}
    Y_{mn} = \sum_{l = 1}^{W} y_{l + (n-1)H - \frac{W}{2}} \, g_l \, e^{-j 2\pi (l-1) (m-1)/N_\mathrm{fft}} \in \mathbb{C},
\end{equation}
where we set $y_l = 0$ whenever $l \leq 0$ or $l > L$ (zero-padding), $n \in \{1, \ldots, N\}$ is the index of the $n$-th time frame and $m \in \{1, \ldots, M\}$ is the index of the $m$-th frequency bin. $N$ is the number of time frames, which can be expressed as a function of $L$, $W$ and $H$. $M \leq N_\mathrm{fft}$ is the number of selected frequency bins. 
\dav{For real-valued signals, the DFT possesses Hermitian symmetry across frequency, therefore in the following we only consider the first $M = N_\mathrm{fft}/2 + 1$ frequency bins without loss of information.}
In general, $N_\mathrm{fft} = W$, which is the minimum value for the existence of an inverse transform. As such, a standard choice is $M = W/2 + 1$, which we refer to as {\em complete sampling}. We define the time and frequency samplings (expressed in seconds and Hertz) by the following sets:
\begin{IEEEeqnarray}{rClcl}
\label{eq:time-frames}
\mathcal{T} &=& \Bigl\{t_n &:=&\; (n-1)\frac{H}{f_s}\Bigr\}_{n=1}^{N},\\
\label{eq:freq-bins}
\mathcal{F} &=& \Bigl\{f_m &:=&\; \frac{m-1}{M-1}\frac{f_s}{2}\Bigr\}_{m=1}^{M}.
\end{IEEEeqnarray}
The t-f support $\mathcal{S}$ is defined as 
\begin{equation}
    \mathcal{S} = \mathcal{F} \times \mathcal{T}.
\label{eq:support}
\end{equation}
In this paper, the STFT will be computed using the Hann analysis window $\mathbf{g} \in \mathbb{R}^W$.
Finally, we define the {\em spectrogram} as the power magnitude of the STFT, denoted as $\mathbf{X} = \vert \mathbf{Y} \vert^2$.

\subsection{Elements of optimal transport}\label{sec:OT}

\dav{In this section, we concisely introduce some important notions of OT that will be necessary for our work, and we refer to \cite{peyre2020computationalot} for a more in-depth introduction to OT.} Let $\alpha = \sum_{i = 1}^{I} a_{i}\delta_{s^\alpha_{i}} \in \mathcal{M}_{+}^{1}(\mathcal{S}_\alpha)$ and $\beta = \sum_{j = 1}^{J} b_{j} \delta_{s^\beta_{j}} \in \mathcal{M}_{+}^{1}(\mathcal{S}_\beta)$ be two probability distributions supported by two possibly distinct supports $\mathcal{S}_\alpha$ and $\mathcal{S}_\beta$. The \textit{optimal transport} between $\alpha$ and $\beta$ is described by the following optimization problem
\begin{equation}
\mathrm{OT}(\alpha, \beta) := \mathrm{OT}_{\mathbf{C}}(\mathbf{a}, \mathbf{b}) = \underset{ \mathbf{T} \in \mathcal{U}(\mathbf{a}, \mathbf{b}) }{ \mathrm{min} }\langle \mathbf{C}, \mathbf{T} \rangle,
\label{eq:ot}
\end{equation}
where $\mathbf{C} \in \mathbb{R}^{I \times J}_+$ is the cost matrix defined by $C_{ij} =c(s^\alpha_{i}, s^\beta_{j})$ with $c$ a cost function (such as the squared Euclidean distance). The feasible set $\mathcal{U}(\mathbf{a},\mathbf{b}) = \{ \mathbf{T} \in \mathbb{R}_{+}^{I \times J} \;|\; \mathbf{T} \mathbf{1}_{J} = \mathbf{a},  \mathbf{T}^\top \mathbf{1}_{I} = \mathbf{b} \}$ enforces the conservation of mass constraint. Matrices $\mathbf{T} \in \mathcal{U}(\mathbf{a},\mathbf{b})$ are called \textit{transport plans}. They indicate how much \textit{mass} $a_i$ is transported from source point $s_i^\alpha$ to target location $s_j^\beta$ in order to obtain the distribution $\beta$. 

Let $\mathcal{S}_\gamma = \{s^\gamma_k\}_{k = 1}^{K} \subset \mathbb{R}^D$ be an arbitrary support, possibly distinct from $\mathcal{S}_\alpha$ and $\mathcal{S}_\beta$. Given a barycentric parameter $\lambda \in [0, 1]$, the \textit{fixed-support OT barycenter} $\gamma_\lambda$ between the probability distributions $\alpha$ and $\beta$ is defined by
\bal{
\gamma_\lambda = \sum_{k = 1}^{K} g_{\lambda, k} \, \delta_{s^\gamma_k} \in \mathcal{M}^{1}_{+}(\mathcal{S}_\gamma)
}
with
\begin{equation}
\mathbf{g}_\lambda \in \underset{\mathbf{g}  \in \mathbb{S}^K}{\mathrm{argmin}} \ (1 - \lambda) \, \mathrm{OT}_{\mathbf{C}_\alpha}(\mathbf{a}, \mathbf{g}) + \lambda \, \mathrm{OT}_{\mathbf{C}_\beta}(\mathbf{b}, \mathbf{g}),
\label{eq:wass-bary}
\end{equation}
where $\mathbf{C}_\alpha$ (resp., $\mathbf{C}_\beta$) defines a cost matrix between $\mathcal{S}_\alpha$ and $\mathcal{S}_\gamma$ (resp., $\mathcal{S}_\beta$ and $\mathcal{S}_\gamma$).

\subsection{Unbalanced optimal transport} \label{section:UOT}

We will see in Section \ref{section:initial-results} that the conservation of mass constraint may produce undesirable effects in our setting. To mitigate this, such constraint can be replaced by a penalization term which allows some of the mass to be left unmatched. This framework is called unbalanced optimal transport (UOT) \cite{séjourné2023unbalancedoptimaltransporttheory}. Given two non-negative distributions $\alpha \in \mathcal{M}_{+}(\mathcal{S}_\alpha)$ and $\beta \in \mathcal{M}_{+}(\mathcal{S}_\beta)$, UOT is described by the following optimization problem
\begin{IEEEeqnarray}{ll}
\label{eq:uot}
\mathrm{UOT}(\alpha,\beta) := \mathrm{UOT}_{\mathbf{C}}(\mathbf{a}, \mathbf{b}) = \\
 \min_{\mathbf{T} \in\mathbb{R}_+^{I\times J}}\ \langle \mathbf{C},\mathbf{T}\rangle \,+\, \eta_1 D(\mathbf{T}\mathbf{1}_J,\mathbf{a}) \nonumber 
    \,+\, \eta_2 D(\mathbf{T}^\top\mathbf{1}_I,\mathbf{b}) 
\end{IEEEeqnarray}
where $\eta_1, \eta_2 > 0$ are the marginal constraint relaxation parameters (or UOT parameters) and $D$ is a measure of fit, here chosen to be the KL divergence \eqref{eq:kl}. Using the notations of Section~\ref{sec:OT}, the \textit{fixed-support UOT barycenter} between the non-negative distributions $\alpha$ and $\beta$ is defined as
\begin{equation}
    \gamma_{\lambda} = \sum_{k = 1}^{K} g_{\lambda,k} \, \delta_{s^\gamma_k} \in \mathcal{M}_{+}(\mathcal{S}_\gamma),
\end{equation}
where
\begin{equation}
\begin{aligned}
\mathbf{g}_\lambda \in \underset{\mathbf{g} \in \mathbb{R}_{+}^K}{\mathrm{argmin}}\,(1 - \lambda) \mathrm{UOT}_{\mathbf{C}_\alpha}(\mathbf{a}, \mathbf{g}) + \lambda \mathrm{UOT}_{\mathbf{C}_\beta}(\mathbf{b}, \mathbf{g}).
\label{eq:wass-bary-uot}
\end{aligned}
\end{equation}
In the UOT literature, \eqref{eq:wass-bary-uot} is generally augmented with an entropic regularization term \cite{chizat2017scalingalgorithmsunbalancedtransport, séjourné2023unbalancedoptimaltransporttheory}. This allows to produce fast Sinkhorn-like algorithms, such as the implementation available in the POT library \cite{POT}. However, entropic regularization tends to produce diffuse barycentric distributions, which is undesirable in our setting where sharp t-f energy localization is desired. Moreover, entropic regularization requires tuning an additional hyper-parameter. 
We will present in Section~\ref{section:algorithm} a novel algorithm that directly solves the optimization problem~\eqref{eq:wass-bary-uot}. To the best of our knowledge, no other existing method solves this problem  without entropic regularization.

\subsection{Spectrograms as non-negative distributions} \label{section:ot-specs}

One key standpoint of our work is to treat t-f distributions (and more precisely, spectrograms) as 
distributions in $\mathcal{M}_{+}(\mathcal{S})$. The t-f grid of the spectrogram is interpreted as the support of a non-negative distribution, while its coefficients are interpreted as the weights that characterize this distribution. Such an interpretation enables OT and UOT computations on spectrograms. This formalism was introduced in our previous conference article \cite{valdivia2025audio}, for a different problem (interpolation of the spectrograms of different sounds supported by the same t-f grid). 

We will next use the following notations. Let $\mathbf{X} \in \mathbb{R}_{+}^{M \times N}$ be a spectrogram with  support $\mathcal{S} \subseteq \mathbb{R}^2$ as defined in \eqref{eq:support}. We may vectorize $\mathbf{X}$ following \eqref{eq:vectorization} to produce a vector $\mathbf{x} \in \mathbb{R}_{+}^{MN}$. Using the relationship $\pi(m, n) = i$ from \eqref{eq:mapping}, we denote $s_{i} = (f_m, t_n)$, leading to $\mathcal{S} = \{s_i\}_{i = 1}^{MN}$. With these notations we associate $\mathbf{X}$ with the non-negative discrete measure
\begin{equation}
    \chi = \sum_{i = 1}^{MN} x_i \delta_{s_{i}} \in \mathcal{M}_{+}(\mathcal{S}).
\label{eq:spectrogram-prob-measure}
\end{equation}
To enable OT computations as in \eqref{eq:ot}, we further require $\chi \in \mathcal{M}_{+}^1(\mathcal{S})$. To do so, we may normalize the spectrogram coefficients so that $\mathbf{x} \in \mathbb{S}^{MN}$. Thus, whenever OT is used in the following, we suppose that this normalization step is performed. This is not required for UOT, which operates on non-negative distributions.

\section{Super-resolution by fusion of spectrograms} \label{section:spec-fusion}

\begin{figure}[t]
\centering
\subfloat[]{\includegraphics[width=\columnwidth]{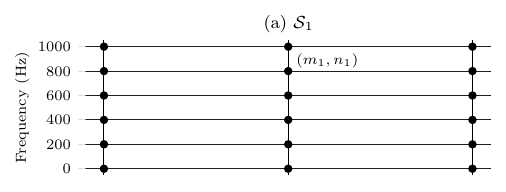}}%
\label{fig:support_high_freq}
\\[-8.5mm]
\subfloat[]{\includegraphics[width=\columnwidth]{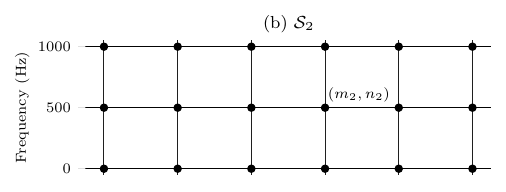}}%
\label{fig:support_high_time}
\\[-8.5mm]
\subfloat[]{\includegraphics[width=\columnwidth]{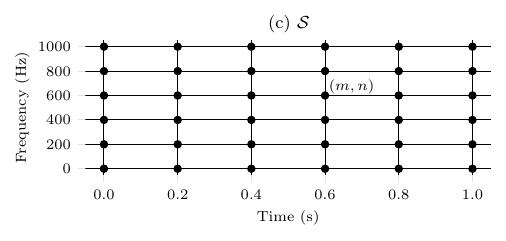}}%
\label{fig:support_high_res}
\\[-11.5mm]
\caption{Time-frequency supports of input and output spectrograms. (a) Support $\mathcal{S}_1 = \mathcal{F}_1\times\mathcal{T}_1$ associated with a long window size. (b) Support $\mathcal{S}_2 = \mathcal{F}_2\times\mathcal{T}_2$ associated with a short window size. (c) Support $\mathcal{S}$ of a canonical super-resolution spectrogram, where frequency samplings are taken from $\mathcal{S}_1$ and time samplings are taken from $\mathcal{S}_2$. In other words, $\mathcal{S} = \mathcal{F}_1\times\mathcal{T}_2 $.
For visual clarity, the high-frequency resolution support $\mathcal{S}_1$ is shown with finer frequency sampling, and the high-temporal resolution support $\mathcal{S}_2$ is shown with finer temporal sampling. 
}
\label{fig:supports}
\end{figure}

\noindent In this section we formulate our approach to super-resolution through the fusion of two spectrograms using OT barycenters. Our approach can also accommodate the fusion of more spectrograms, and this will be discussed in Section~\ref{sec:multi}. Let $\mathbf{X}_1 \in \mathbb{R}_{+}^{M_1 \times N_1}$ and $\mathbf{X}_2 \in \mathbb{R}_{+}^{M_2 \times N_2}$ be two spectrograms of the same signal, computed with analysis windows of different lengths. We suppose that $\mathbf{X}_1$ is computed with a long window (high frequency resolution) and $\mathbf{X}_2$ with a short window (high temporal resolution). Let $\mathcal{S}_1$ and $\mathcal{S}_2$ denote their respective t-f supports. Let also $\mathcal{S}$ denote the user-defined t-f support for the super-resolution spectrogram. In this paper we assume that $\mathcal{S}$ is a t-f grid (of arbitrary size $M \times N$), though our approach could more generally accommodate irregular t-f supports. A canonical choice for $\mathcal{S}$ is presented in Fig.~\ref{fig:supports}.

\subsection{Cost of transportation in the time-frequency plane}
Given arbitrary supports ${\cal S}^\alpha, {\cal S}^\beta \subset \mathbb{R}^D$ and $(s_{i}^\alpha, s_{j}^\beta) \in {\cal S}^\alpha \times {\cal S}^\beta $, a standard transportation cost function in OT is the squared Euclidean distance, given by
\begin{equation}
c(s_{i}^\alpha, s_{j}^\beta)=\Vert s_{i}^\alpha - s_{j}^\beta\Vert^2.
\end{equation}
Such a choice produces a Wasserstein barycenter \cite{peyre2020computationalot}. However, this choice is not directly appropriate for t-f supports, since time and frequency are heterogeneous quantities and should not be compared on the same scale. We therefore compare t-f points through normalized grid coordinates. Using the mapping $\pi(m,n)=i$ from \eqref{eq:mapping}, we associate a point $s_i = (f_m,t_n) \in \mathcal{S}$ to the normalized coordinate pair 

\begin{equation}
    \biggl(\frac{m-1}{M-1},\frac{n-1}{N-1}\biggr),
\end{equation}
and similarly for $\mathcal{S}_1$ and $\mathcal{S}_2$. With this, we define the cost functions $c_1$ and $c_2$ as
\begin{IEEEeqnarray}{rCl}
c_1(s_{i_1},s_{i}) &=& \left(\frac{m_1-1}{M_1-1}-\frac{m-1}{M-1}\right)^2
+
\left(\frac{n_1-1}{N_1-1}-\frac{n-1}{N-1}\right)^2, \nonumber\\
c_2(s_{i_2},s_{i}) &=& \left(\frac{m_2-1}{M_2-1}-\frac{m-1}{M-1}\right)^2
+
\left(\frac{n_2-1}{N_2-1}-\frac{n-1}{N-1}\right)^2, \nonumber\\
\end{IEEEeqnarray}
where $s_{i_1} \in \mathcal{S}_1$, $s_{i_2} \in \mathcal{S}_2$ and $s_i \in \mathcal{S}$.
\footnote{\dav{For conciseness, we are slightly abusing the notations by merely distinguishing the points $s_{i_1} \in {\cal S}_1 $, $s_{i_2} \in {\cal S}_2$ and $s_{i} \in {\cal S}$ by the notational convention for their indices, namely $i_1$, $i_2$ and $i$.
}} 
These cost functions define the cost matrices $\mathbf{C}_1 \in \mathbb{R}_{+}^{M_1N_1 \times MN}$ and $\mathbf{C}_2 \in \mathbb{R}_{+}^{M_2N_2 \times MN}$ with entries given by:
\begin{equation}
C_{1,i_1i} = c_1(s_{i_1},s_{i}), \qquad (s_{i_1},s_{i}) \in \mathcal{S}_1 \times \mathcal{S},
\label{eq:c1}
\end{equation}
\begin{equation}
C_{2,i_2i} = c_2(s_{i_2},s_{i}), \qquad (s_{i_2},s_{i}) \in \mathcal{S}_2 \times \mathcal{S}.
\label{eq:c2}
\end{equation}

\subsection{OT-based spectrogram fusion} \label{section:ot-based-spectrogram-fusion}
Let $\chi_1 \in \mathcal{M}_{+}^1(\mathcal{S}_1)$ and $\chi_2 \in \mathcal{M}_{+}^1(\mathcal{S}_2)$ denote the probability distributions associated with $\mathbf{X}_1$ and $\mathbf{X}_2$ (following Section \ref{section:ot-specs}), with normalized weights $\mathbf{x}_1 \in \mathbb{S}^{M_1 N_1}$ and $\mathbf{x}_2 \in \mathbb{S}^{M_2 N_2}$, respectively. We define the fusion of spectrograms $\mathbf{X}_1$ and $\mathbf{X}_2$ as the OT barycenter 
\begin{equation}
    \chi = \sum_{i = 1}^{MN} x_i \delta_{s_i} \in \mathcal{M}_{+}^1(\mathcal{S}),
    \label{eq:ot-barydef}
\end{equation}
whose weights are given by
\begin{equation}
    \mathbf{x} \in \underset{\mathbf{x} \in \mathbb{S}^{MN}}{\mathrm{argmin}}\; (1 - \lambda) \, \mathrm{OT}_{\mathbf{C}_1}(\mathbf{x}_1, \mathbf{x}) + \lambda \, \mathrm{OT}_{\mathbf{C}_2}(\mathbf{x}_2, \mathbf{x}).
\label{eq:ot-bary}
\end{equation}
The super-resolution spectrogram $\mathbf{X} \in \mathbb{R}_{+}^{M \times N}$ can then be retrieved from $\mathbf{x}$ by inverting the vectorization \eqref{eq:vectorization}. By construction, $\mathbf{X}$ has unit global energy. {Note that} when necessary, it may be rescaled, for example by using the barycentric energy of the input spectrograms. 
\dav{In the remainder of the paper, and without loss of generality, we set \(\lambda=0.5\), assuming that the two spectrograms \(\mathbf{X}_1\) and \(\mathbf{X}_2\) provide equally important information.}

\subsection{Preliminary results} \label{section:initial-results}

\begin{figure*}[t]
\centering
\subfloat[]{\includegraphics[width=0.5\columnwidth]{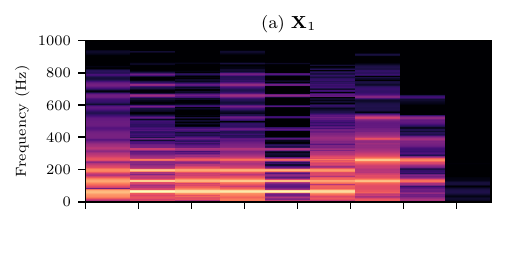}}%
\label{fig:fusion-large}
\hspace{-4mm}
\subfloat[]{\includegraphics[width=0.5\columnwidth]{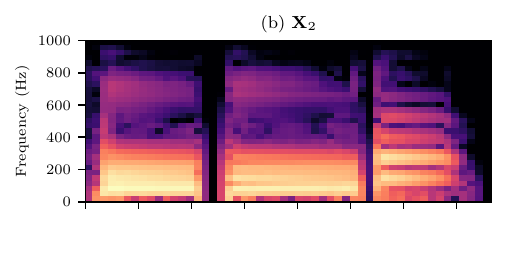}}%
\label{fig:fusion-short}
\\[-10mm]
\subfloat[]{\includegraphics[width=0.5\columnwidth]{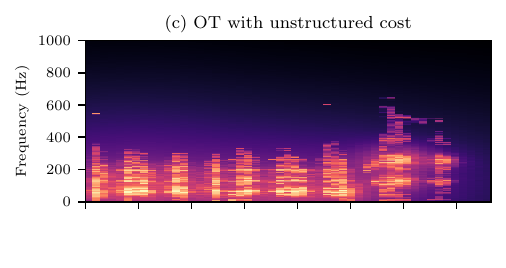}}%
\label{fig:fusion-ot}
\hspace{-4mm}
\subfloat[]{\includegraphics[width=0.5\columnwidth]{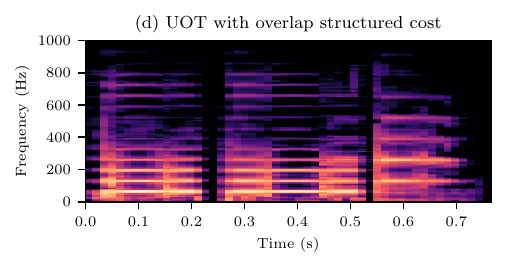}}%
\label{fig:fusion-uot}
\\[-10mm]
\caption{Fusion results for a signal consisting of three bass-guitar notes. Amplitude is displayed on a logarithmic scale. (a) Spectrogram $\mathbf{X}_1$ computed with a long window ($200\,\mathrm{ms}$). (b) Spectrogram $\mathbf{X}_2$ computed with a short window ($30\,\mathrm{ms}$). (c) OT barycenter obtained with unstructured squared Euclidean cost matrices \eqref{eq:c1}-\eqref{eq:c2}. (d) UOT barycenter obtained with the structured cost matrices \eqref{eq:cost-overlap-time}-\eqref{eq:cost-overlap-freq}, with \(\eta = 1\).}
\label{fig:fusion}
\end{figure*}

We present an initial set of results that illustrate both the potential and the limitations of OT-based fusion. Our analysis focuses on a musical signal consisting of three bass-guitar notes, with a duration of 800~ms and a sampling frequency of 2~kHz. The spectrogram $\ve{X}_1$ is computed using a long window of 200~ms, while $\ve{X}_2$ is computed using a short window of 30~ms. The hop size is set to half the window length and $M$ is set to complete sampling.
The resulting spectrograms are shown in Fig.~\ref{fig:fusion}~(a)--(b). Following the setting illustrated in Fig.~\ref{fig:supports}, we consider the canonical target support defined as
\begin{equation}
    \mathcal{S} = \mathcal{F}_1 \times \mathcal{T}_2,
\end{equation}
where $\mathcal{F}_1$ is the frequency sampling of $\mathbf{X}_1$ and $\mathcal{T}_2$ is the time sampling of $\mathbf{X}_2$. This configuration yields $M = M_1$ and $N = N_2$. The barycenter~\eqref{eq:ot-barydef} of the spectrograms is computed \dav{by running 1000 iterations of} the algorithm proposed in~\cite{cuturi2014fast}. The resulting super-resolution spectrogram is shown in Fig.~\ref{fig:fusion}~(c). While the fusion is effective, it fails to fully preserve the expected t-f structure: energy is dispersed across t-f points, and harmonics above 400~Hz are strongly attenuated. Moreover, the computation of the OT barycenter is expensive, both in time and memory. 

Indeed, with $\mathcal{S} = \mathcal{F}_1 \times \mathcal{T}_2$, the cost matrices $\ve{C}_1$ and $\ve{C}_2$ contain $M_1^2 N_1 N_2 = 1.9\times 10^{7}$ and $M_1 M_2 N_2^2 = 1.6 \times 10^{7}$ entries, respectively, \dav{as reported in Table~\ref{tab:cost_matrices}}. On a laptop equipped with an Apple M4 chip and 24~GB of RAM, the computation of $\ve{X}$ took about $32$ minutes, despite the signal duration being less than one second and its sampling frequency being only 2~kHz. In the following section, we address these two limitations—precision and computational efficiency—by introducing structured cost matrices in the UOT setting.

\begin{table}[t]
\centering
\caption{\dav{Comparison of the cost matrices and barycenter computation times for the example presented in Fig. \ref{fig:fusion}.}}
\label{tab:cost_matrices}
\small
\renewcommand{\arraystretch}{1.2}
\setlength{\tabcolsep}{7pt}
\begin{tabular}{lccc}
\toprule
& Unstructured & Axis-only & Overlap  \\
\midrule
Section
& III-C & III-D & III-E \\


Method
& OT & UOT & UOT \\

Equations
& \eqref{eq:c1}-\eqref{eq:c2} & \eqref{eq:cost-matrix-inf-f}-\eqref{eq:cost-matrix-inf-t} & \eqref{eq:cost-overlap-time}-\eqref{eq:cost-overlap-freq}  \\

Cost matrices
& $\mathbf{C}_1, \mathbf{C}_2$
& $\bar{\mathbf{C}}_1, \bar{\mathbf{C}}_2$
& $\tilde{\mathbf{C}}_1, \tilde{\mathbf{C}}_2$ \\

\midrule
\multicolumn{4}{l}{\textit{Nb of finite entries}} \\

Cost $\mathcal{S}_1 \times \mathcal{S}$
& $1.9\times10^7$
& $9.4\times10^4$
& $2.4\times10^4$ \\

Cost $\mathcal{S}_2 \times \mathcal{S}$
& $1.7\times10^7$
& $3.2\times10^5$
& $4.8\times10^4$ \\

\midrule
Runtime (s) & $1945$ & $30$ & $5$  \\
\bottomrule
\end{tabular}
\end{table}

\subsection{Structured cost matrices and UOT} \label{section:structured-cost-matrix}

The squared Euclidean cost matrices $\mathbf{C}_1$ and $\mathbf{C}_2$ defined by~\eqref{eq:c1}-\eqref{eq:c2} allow t-f energy to move freely across the target t-f support, with the undesirable effects illustrated in Fig.~\ref{fig:fusion}~(c). Given that $\mathbf{X}_1$ already ensures accurate frequency localization, its energy should be redistributed exclusively along the time axis. Conversely, since $\mathbf{X}_2$ provides accurate time localization, its energy should be redistributed solely along the frequency axis. With the target support $\mathcal{S}=\mathcal{F}_1\times\mathcal{T}_2$, we may enforce these requirements with the cost matrices $\bar{\mathbf{C}}_1\in\mathbb{R}^{M_1N_1\times MN}$ and $\bar{\mathbf{C}}_2\in\mathbb{R}^{M_2N_2\times MN}$, defined entrywise by
\begin{equation}
\bar{C}_{1,i_1 i}
=
\begin{cases}
c_1(s_{i_1},s_i), & \text{if } m_1=m,\\
+\infty, & \text{otherwise,}
\end{cases}
\label{eq:cost-matrix-inf-f}
\end{equation}
\begin{equation}
\bar{C}_{2,i_2 i}
=
\begin{cases}
c_2(s_{i_2},s_i), & \text{if } n_2=n,\\
+\infty, & \text{otherwise,}
\end{cases}
\label{eq:cost-matrix-inf-t}
\end{equation}
where we recall that $i_1 = \pi(m_1,n_1)$, $i_2 = \pi(m_2,n_2)$ and $i = \pi(m,n)$.

Note that setting ${C}_{ij} = + \infty$ in \eqref{eq:ot} implicitly enforces the transport plan to verify $T_{ij} = 0$. With standard OT, such hard constraints may lead to infeasibility under mass conservation, whereas the unbalanced formulation in \eqref{eq:uot} remains well-defined \cite{peyre2020computationalot}.

\dav{In the example introduced in Section~\ref{section:initial-results}, enforcing the constraint in~\eqref{eq:cost-matrix-inf-f}-\eqref{eq:cost-matrix-inf-t} results in reducing the number of finite entries in  $\bar{\ve{C}}_1$ and $\bar{\ve{C}}_2$ to $9.2\times 10^{4}$ and $3.2\times 10^{5}$, respectively (see  Table~\ref{tab:cost_matrices}).} We computed the UOT barycenter of the unnormalized spectrograms $\ve{X}_1$ and $\ve{X}_2$ (corresponding to \eqref{eq:ot-bary} where OT is replaced with UOT) \dav{by running 1000 iterations of} the algorithm presented in Section~\ref{section:algorithm}. As discussed in Section~\ref{section:algorithm}, our algorithm can efficiently ignore the infinite entries in $\bar{\ve{C}}_1$ and $\bar{\ve{C}}_2$. \dav{Table~\ref{tab:cost_matrices} shows that this considerably improves computational efficiency}.

Additionally, unlike OT, UOT does not require normalization of the input spectrograms. This allows the barycenter to be computed directly from the original spectrogram energies, with its total mass determined automatically during optimization, eliminating the need for post-processing normalization.

\subsection{Further refinement of the cost matrices} \label{section:overlap-cost-matrix}

\begin{figure}
    \centering
    \includegraphics[width=\columnwidth]{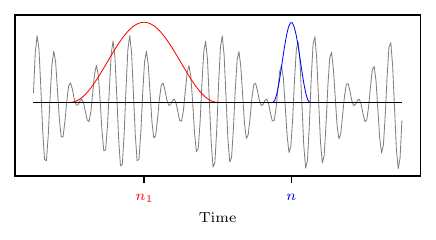}
    \\[-4mm]
    \caption{Signal windowing with windows of different lengths. The long window (in red) corresponds to frame $t_{n_1}$ in $\mathbf{X}_1$. The short window (in blue) corresponds to frame $t_{n}$ in $\mathbf{X}$. In this case, windows do not overlap in the time domain, so we forbid 
    transportation of mass from frame $t_{n_1}$ to frame $t_{n}$.}
    \label{fig:temporal-overlap}
\end{figure}

The structured matrices $\bar{\ve{C}}_1$ and $\bar{\ve{C}}_2$ still allow transport between any pair of aligned time or frequency samplings, {respectively}, which remains too permissive. Consider the transport from $\mathbf{X}_1$ to $\mathbf{X}$. At a given frequency $f_m$, energy may be moved from a frame $t_{n_1}\in\mathcal{T}_1$ to any frame $t_{n}\in\mathcal{T} = \mathcal{T}_2 $, even when the corresponding analysis windows do not overlap in time. Such displacements are implausible because non-overlapping windows do not capture common signal content, as illustrated by Fig.~\ref{fig:temporal-overlap}. The same reasoning applies to energy displacements along frequencies at a given frame, in the transport from $\mathbf{X}_2$ to $\mathbf{X}$.
Therefore, we incorporate supplementary overlap constraints in the design of the cost matrices, in order to restrict energy displacements to neighboring frames in $\bar{\ve{C}}_1$, and to neighboring frequencies in $\bar{\ve{C}}_2$. These constraints increase the number of infinite entries in the cost matrices, which further accelerates computations. 

More precisely, let $\mathcal{O}_1(n_1)$ be the set of indices $n$ corresponding to frames $t_{n}$ of $\mathbf{X}$ that overlap with frame $t_{n_1}$ of $\mathbf{X}_1$ (the set of ``neighboring'' time frames, see the Supplementary Material for a rigorous definition). We replace the cost matrix $\mathbf{\bar{C}}_1$ from \eqref{eq:cost-matrix-inf-f} by the cost matrix $\tilde{\mathbf{C}}_1$ defined entrywise as
\begin{equation}
\tilde{C}_{1,i_1i}=
\begin{cases}
c_1(s_{i_1}, s_{i}), & \text{if } m_1=m \text{ and } n\in\mathcal{O}_1(n_1),\\
+\infty, & \text{otherwise.}
\end{cases}
\label{eq:cost-overlap-time}
\end{equation}
Similarly, let $\mathcal{O}_2(m_2)$ denote the set of indices $m$ such that the effective frequency support induced by the analysis window at frequency $f_m$ in $\mathbf{X}$ overlaps with the one induced by the analysis window at frequency $f_{m_2}$ in $\mathbf{X}_2$, that is, the set of neighboring frequency bins (see the Supplementary Material for a rigorous definition).
We replace $\mathbf{\bar{C}}_2$ from \eqref{eq:cost-matrix-inf-t} by the cost matrix $\tilde{\mathbf{C}}_2$ defined entrywise as
\begin{equation}
\tilde{C}_{2,i_2i}=
\begin{cases}
c_2(s_{i_2}, s_{i}), & \text{if } n_2=n \text{ and } m\in\mathcal{O}_2(m_2),\\
+\infty, & \text{otherwise.}
\end{cases}
\label{eq:cost-overlap-freq}
\end{equation}
\dav{In the experimental setting of Section \ref{section:initial-results}, this reduces even further the number of finite entries and computation times, see Table~\ref{tab:cost_matrices}. The resulting barycenter, shown in Fig.~\ref{fig:fusion}(d), effectively captures the localization properties of the input spectrograms, clearly improving the fusion quality with respect to the standard OT-based approach with the unstructured cost matrices \eqref{eq:c1}-\eqref{eq:c2}. We also computed the UOT barycenter with the axis-based cost matrices \eqref{eq:cost-matrix-inf-f}-\eqref{eq:cost-matrix-inf-t} and the resulting spectrogram is  visually nearly identical to the one obtained with cost matrices \eqref{eq:cost-overlap-time}-\eqref{eq:cost-overlap-freq}. This similarity is further demonstrated quantitatively in Section~\ref{sec:synth-signals}.}

\section{Algorithm}\label{section:algorithm}

\noindent In this section, we propose a novel block majorization-minimization (MM) algorithm to solve the UOT barycenter problem \eqref{eq:wass-bary-uot} when the input distributions are defined on different supports. To the best of our knowledge, no other algorithm addresses this specific problem in the current literature.

\subsection{Reformulation of the objective function} \label{sec:obj}
We aim at minimizing the objective function that appears in the definition of the fixed-support UOT barycenter \eqref{eq:wass-bary-uot}. With \eqref{eq:uot}, this function can be written as
\begin{IEEEeqnarray}{l} \label{eq:obj}
    Q_\lambda(\mathbf{g}) = \nonumber\\
    \quad(1 - \lambda) \underset{\mathbf{T}_\alpha \in \mathbb{R}_{+}^{I \times K}}{\mathrm{min}} U_\alpha(\mathbf{T}_\alpha, \mathbf{g})  + \lambda \,\underset{\mathbf{T}_\beta \in \mathbb{R}_{+}^{J \times K}}{\mathrm{min}} U_\beta(\mathbf{T}_\beta, \mathbf{g}), \nonumber\\
\end{IEEEeqnarray}
where 
\begin{IEEEeqnarray}{lll}
\label{eq:f_alpha}
U_\alpha(\mathbf{T}_\alpha, \mathbf{g}) =&\; \langle \mathbf{C}_\alpha, \mathbf{T}_\alpha \rangle &\, + \,\eta^\alpha_1 \, \mathrm{KL}(\mathbf{T}_\alpha \mathbf{1}_K, \mathbf{a}) \nonumber \\ &&\, + \, \eta^\alpha_2 \, \mathrm{KL}(\mathbf{T}_\alpha^\top \mathbf{1}_I, \mathbf{g}),  \\
\label{eq:f_beta}
U_\beta(\mathbf{T}_\beta, \mathbf{g}) =&\; \langle \mathbf{C}_\beta, \mathbf{T}_\beta \rangle & \,+ \, \eta^\beta_1 \, \mathrm{KL}(\mathbf{T}_\beta \mathbf{1}_K, \mathbf{b}) \nonumber \\ &&\, + \, \eta^\beta_2 \, \mathrm{KL}(\mathbf{T}_\beta^\top \mathbf{1}_J, \mathbf{g}).
\end{IEEEeqnarray}
Given that $\mathbf{T}_\alpha$ and $\mathbf{T}_\beta$ are independent variables, we can factor out the $\min$ operators in \eqref{eq:obj} and write
\begin{equation}
    Q_\lambda(\mathbf{g}) = \underset{\substack{
        \mathbf{T}_\alpha  \in \mathbb{R}^{I \times K}_{+} ,   \\
    \mathbf{T}_\beta   \in \mathbb{R}^{J \times K}_{+} }}{\mathrm{min}} F_\lambda(\mathbf{T}_\alpha, \mathbf{T}_\beta, \mathbf{g}),
\end{equation}
with 
\begin{equation}
    F_\lambda(\mathbf{T}_\alpha, \mathbf{T}_\beta, \mathbf{g}) = (1 - \lambda) U_\alpha(\mathbf{T}_\alpha, \mathbf{g}) + \lambda U_\beta(\mathbf{T}_\beta, \mathbf{g}).
\label{eq:objective}
\end{equation}
As a consequence, solving \eqref{eq:wass-bary-uot} is equivalent to minimizing $F_\lambda(\mathbf{T}_\alpha, \mathbf{T}_\beta, \mathbf{g})$ with respect to its three variables. \dav{Note that thanks to the joint-convexity of the KL divergence, $F_\lambda$ is a convex function, because it consists of sums and compositions of convex terms with affine mappings.}

\subsection{Block MM algorithm} \label{sec:BMM}
We propose a block-descent algorithm for the minimization of $F_\lambda$ based on the alternate minimization over the three blocks $\mathbf{T}_\alpha$, $\mathbf{T}_\beta$ and $\mathbf{g}$, and resulting in Algorithm~\ref{alg1}.

\subsubsection{{Updates of $\mathbf{T}_\alpha$ and $\mathbf{T}_\beta$}}

Given $\mathbf{g}$ and $\mathbf{T}_\beta$, the minimization of $F_\lambda$ w.r.t $\mathbf{T}_\alpha$ boils down to the minimization of $U_\alpha(\mathbf{T}_\alpha,\ve{g})$ w.r.t $\mathbf{T}_\alpha$. Therefore, the majorization-minimization (MM) update proposed in \cite{chapel2021unbalanced} can be used for updating $\mathbf{T}_\alpha$, and is given in Algorithm \ref{alg1}. A similar procedure is used to update $\mathbf{T}_\beta$ for fixed $\mathbf{g}$ and $\mathbf{T}_\alpha$.

\subsubsection{{Update of $\mathbf{g}$}}
Given $\mathbf{T}_\alpha$ and $\mathbf{T}_\beta$, the minimization of $F_\lambda$ w.r.t $\ve{g}$ boils down to 
\bal{
\underset{\mathbf{g} \in \mathbb{R}_{+}^K}{\mathrm{min}}
    (1 - \lambda)\eta^\alpha_2\,\mathrm{KL}(\mathbf{T}_\alpha^\top \mathbf{1}_I, \mathbf{g})
    + \lambda \eta^\beta_2\,\mathrm{KL}(\mathbf{T}_\beta^\top \mathbf{1}_J, \mathbf{g}).
}
The minimizer admits a closed form, given in Algorithm \ref{alg1}.

\subsubsection{{Summary}}
Our algorithm for solving~\eqref{eq:wass-bary-uot} is summarized in Algorithm \ref{alg1}. The variables $\mathbf{T}_\alpha$, $\mathbf{T}_\beta$, and $\mathbf{g}$ are initialized with ones and are updated until a convergence criterion is met. By construction, our block-descent MM algorithm ensures the monotonicity of the values of $F_\lambda$ at every iteration. Given that $F_\lambda $ is bounded below by zero, the objective values converge. 
\dav{Our algorithm is a special case of the general BSUM framework (Block Successive Upper-bound Minimization) introduced in \cite{razaviyayn2013unified}. However our problem does not satisfy the technical conditions presented in this paper to state the convergence of the iterates to a critical point, which is left for future work. In the experiments of Sections~\ref{sec:synth-signals} and~\ref{sec:real-signals}, we stop the iterations when the following criterion is met:
\begin{equation}
    \Biggl\vert\frac{F_\lambda(\theta^{(k)})-F_\lambda(\theta^{(k-1)})}{F_\lambda(\theta^{(k-1)})}\Biggr\vert<10^{-4},
\label{eq:convergence-criterion}
\end{equation}
where $\theta^{(k)} = (\mathbf{T}_\alpha^{(k)}, \mathbf{T}_\beta^{(k)}, \mathbf{g}^{(k)})$.}

\begin{algorithm}[t]
\caption{Block MM for computation of the UOT barycenter of distributions with different supports}
\label{alg1}
\begin{algorithmic}
\STATE {\textbf{Input}:} Non-negative weights $(\mathbf{a,b})$, barycenter parameter $\lambda \in [0, 1]$, cost matrices $(\mathbf{C}_\alpha, \mathbf{C}_\beta)$, UOT parameters $\eta^\alpha_1, \eta^\alpha_2, \eta^\beta_1, \eta^\beta_2 >0$.
\STATE {\textbf{Output}:} Barycenter weights $\mathbf{g}$, transport plans $(\mathbf{T}_\alpha, \mathbf{T}_\beta)$.
\STATE {\textbf{Initialization}:} 
\STATE \hspace{0.5cm} $\mathbf{g}^0 \gets \mathbf{1}_K, \quad\mathbf{T}^0_\alpha \gets \mathbf{1}_{I \times K}, \quad \mathbf{T}^0_\beta \gets \mathbf{1}_{J \times K}$
\STATE {\textbf{Set}:} 
\STATE \hspace{0.5cm}  $\eta^\alpha \gets \eta^\alpha_1 + \eta^\alpha_2, \eta_\beta \gets \eta^\beta_1 + \eta^\beta_2$.
\STATE \hspace{0.5cm}  $\mathbf{D}_\alpha \gets \exp(-\mathbf{C}_\alpha / \eta^\alpha), \mathbf{D}_\beta \gets \exp(-\mathbf{C}_\beta / \eta_\beta)$ 
\STATE \textbf{Repeat until convergence}:
\STATE \hspace{0.5cm} \textbullet \; Update $\mathbf{T}_\alpha$ and $\mathbf{T}_\beta$ \dav{using \cite{chapel2021unbalanced}}:
\begin{IEEEeqnarray}{l}
    \mathbf{T}^{(k)}_\alpha = \nonumber\\ 
\hspace*{-2mm}    \mathrm{diag}\Biggl(\dfrac{\mathbf{a}}{\mathbf{T}_\alpha^{(k-1)} \mathbf{1}_K}\Biggr)^{\frac{\eta^\alpha_1}{\eta^\alpha}} \Biggl(\mathbf{T}_\alpha^{(k-1)} \odot \mathbf{D}_\alpha \Biggl) \mathrm{diag}\Biggl(\dfrac{\mathbf{g}^{(k-1)}}{\mathbf{T}_\alpha^{(k-1)\top} \mathbf{1}_I}\Biggr)^{\frac{\eta^\alpha_2}{\eta^\alpha}}\nonumber
\end{IEEEeqnarray}
\begin{IEEEeqnarray}{l}
    \mathbf{T}^{(k)}_\beta = \nonumber\\
\hspace*{-2mm}    \mathrm{diag}\Biggl(\dfrac{\mathbf{b}}{\mathbf{T}_\beta^{(k-1)} \mathbf{1}_K}\Biggr)^{\frac{\eta^\beta_1}{\eta_\beta}} \Biggl(\mathbf{T}_\beta^{(k-1)} \odot \mathbf{D}_\beta \Biggl) \mathrm{diag}\Biggl(\dfrac{\mathbf{g}^{(k-1)}}{\mathbf{T}_\beta^{(k-1)\top} \mathbf{1}_J}\Biggr)^{\frac{\eta^\beta_2}{\eta_\beta}}\nonumber
\end{IEEEeqnarray}
\vspace{0.2cm}
\STATE \hspace{0.5cm} \textbullet \; Update $\mathbf{g}$ :
\begin{equation}
    \mathbf{g}^{(k)} = \frac{(1 - \lambda) \, \eta^\alpha_2 \, \mathbf{T}_\alpha^{{(k)}\top} \mathbf{1}_I + \lambda \, \eta^\beta_2 \, \mathbf{T}_\beta^{{(k)}\top} \mathbf{1}_J)}{(1 - \lambda) \, \eta^\alpha_2 \, + \lambda \eta^\beta_2}\nonumber
\end{equation}
\STATE \textbf{Return} $\mathbf{g}^{(k)}, \mathbf{T}_\alpha^{(k)}, \mathbf{T}_\beta^{(k)}$
\end{algorithmic}
\end{algorithm}

\subsection{Exploiting sparsity} \label{section:sparsity}

When computing either $\mathbf{T}_\alpha^{(k)}$ or $\mathbf{T}_\beta^{(k)}$ in Algorithm~\ref{alg1}, each entry (denoted $T^{(k)}_{ij}$ for simplicity) is obtained by multiplying $T^{(k-1)}_{ij}$ with $\exp(-C_{ij}/\eta)$. Thus, when $C_{ij} = +\infty$ then $\exp(-C_{ij}/\eta) = 0$, leading to $T_{ij}^{(k)} = 0$.
 Such entries therefore never contribute to the objective and can be removed from the optimization variables. In practice, this results in a substantial reduction in memory and computational cost. For the cost matrices $\tilde{\ve{C}}_1$ and $\tilde{\ve{C}}_2$ in \eqref{eq:cost-overlap-time}-\eqref{eq:cost-overlap-freq} that incorporate overlap constraints, the number of finite entries is governed by the STFT parameters (window length, hop size, frequency spacing). For example, in the setting of Section \ref{sec:real-signals} involving speech signals, the ratio of finite entries to the total number of entries is approximately $10^{-4}$. \dav{In the Supplementary Material, we provide a complexity analysis of Algorithm~\ref{alg1} under various settings. In particular, we show that the complexity of the algorithm scales linearly with signal length when processing spectrograms with the structured cost matrices $\tilde{\ve{C}}_1$, $\tilde{\ve{C}}_2$ (and standard values of $H$ and $M$), whereas it scales quadratically with the signal length in the general unstructured setting.}

\subsection{Extension to more than two distributions} \label{sec:multi}

So far we have considered the problem of computing the UOT barycenter of two distributions with possibly different supports. We may easily consider the UOT barycenter of more distributions. Given the distributions $\alpha_p \in \mathcal{M}_{+}(\mathcal{S}_{\alpha_p})$, $p=1,\ldots,P$, the arbitrary barycentric support ${\cal S_\gamma}$ and the weights $\boldsymbol{\lambda} \in \mathbb{S}^P$, the barycenter writes
\begin{equation}
    \gamma_{\boldsymbol{\lambda}} = \sum_{k = 1}^{K} g_{\boldsymbol{\lambda},k} \, \delta_{s^\gamma_k} \in \mathcal{M}_{+}(\mathcal{S}_\gamma),
\end{equation}
where
\bal{
\mathbf{g}_{\boldsymbol{\lambda}} \in \underset{\mathbf{g} \in \mathbb{R}_{+}^K}{\mathrm{argmin}}\, \sum_{p=1}^P \lambda_p \mathrm{UOT}_{\mathbf{C}_p}(\mathbf{a}_p, \mathbf{g}),
\label{eq:wass-bary-uot-multi}
}
and where, for $p=1,\ldots, P$, 
\begin{itemize}
    \item the nonnegative vector $\mathbf{a}_p$ denotes the weights of the distribution $\alpha_p$,
    \item the nonnegative matrix $\mathbf{C}_p$ denotes the transportation cost between ${\cal S}_{\alpha_p}$ and ${\cal S}$.
\end{itemize}
We may follow the steps described in Section~\ref{sec:obj} to rewrite problem~\eqref{eq:wass-bary-uot-multi} as the optimization of a function $F_{\boldsymbol{\lambda}}$ composed of $P$ terms involving $\ve{g}$ and the transport plans $\ve{T}_{\alpha_p}$, $p=1,\ldots,P$. From there, we may follow Section~\ref{sec:BMM} to derive a block-descent MM algorithm that involves alternate minimization steps over the $P+1$ variables. Comments on how to produce structured cost matrices in the t-f setting for the fusion of more than two spectrograms are given in the Supplementary Material.

\section{Experiments with synthetic signals} \label{sec:synth-signals}

\noindent In this section, we evaluate the t-f localization properties of the proposed fusion method on controlled synthetic signals. To this end, we generate signals composed of either a single sinusoidal packet (a short localized sine signal) or a mixture of sinusoidal packets with random durations and frequencies. In Section \ref{sec:tf-localization-single}, we first consider a single-packet setting, in which temporal and frequency localization can be assessed separately. In Section \ref{sec:tf-localization-multi}, we analyze mixtures of sinusoidal packets and evaluate overall t-f localization. \dav{Furthermore, we evaluate if the energy is uniformly and smoothly spread within the true t-f support. The corresponding error metrics are introduced in each subsection. We compare our method with the results obtained from 1) raw over-complete input spectrograms, 2) these reassigned input spectrograms \cite{flandrin2018time}, 3) geometric-mean fusion \cite{cheung1991combined}, and 4) NMF-based fusion \cite{Leplat2022}}.

\subsection{Implementation details} \label{sec:imp-details}

\subsubsection{\dav{{Signal description}}}
We run every experiment with $N_\mathrm{sig}=100$ random signals composed as either one or a mixture of sinusoid t-f packets with random frequency and random onset and offset times. The signals are $0.5\,\mathrm{s}$ long and are sampled at $f_s=1\,\mathrm{kHz}$. Outside the support of the packets, the signals are identically zero. 

To generate a random packet, we uniformly sample a frequency $f^\ast \in [200, 400]$, in Hertz, and a duration $d \in [0.01, 0.04]$, in seconds. This range was chosen so that the packets remain short enough to make temporal localization discriminative, while still containing enough  oscillations to retain a meaningful frequency content over the considered frequency range.
The onset time $t^\mathrm{on}$ is uniformly sampled from $[0, 0.5 - d]$ and the offset time is given by $t^\mathrm{off} = t^\mathrm{on} + d$.

\subsubsection{\dav{{UOT hyperparameters}}}

For simplicity, in all experiments, the UOT marginal relaxation parameters of \eqref{eq:objective} are set to be equal, i.e., 
\begin{equation}
  \eta^\alpha_1 =\eta^\alpha_2 = \eta^\beta_1 = \eta^\beta_2 = \eta.
\end{equation}
Empirically, we observe that very small values of $\eta$ lead to excessively sparse transport plans, while large values tend to produce overly diffuse fusion spectrograms. \dav{We choose $\eta=1$ as a compromise between these behaviors. A sensitivity analysis with respect to \(\eta\) is provided in the Supplementary Material.}

\subsubsection{\dav{Experimental setting}}

\dav{We first consider a {\em different-grid setting} in which the input spectrograms do not share the same t-f support. We compute $\mathbf{X}_1\in\mathbb{R}_+^{M_1\times N_1}$ and $\mathbf{X}_2\in\mathbb{R}_+^{M_2\times N_2}$ with window sizes $W_1=100\,\mathrm{ms}$ and $W_2=20\,\mathrm{ms}$. The spectrogram $\mathbf{X}_1$ is computed with frequency spacing $2\,\mathrm{Hz}$ and $75\,\%$ overlap, corresponding to a hop size of $25\,\mathrm{ms}$, whereas $\mathbf{X}_2$ is computed with hop size $2\,\mathrm{ms}$ and complete frequency sampling, corresponding to a frequency spacing of $50\,\mathrm{Hz}$.} \dav{In this scenario we compute the different-grid UOT barycenter with shape $M\times N$, where $M=M_1=257$ and $N=N_2=255$. We also compute three spectrograms fused with the NMF method of \cite{Leplat2022}, using the KL divergence as the measure of fit ($\beta=1$) and varying rank \(K \in \{1, 10, 100\}\), and with all the other hyper-parameters set as in \cite{Leplat2022}.
}

\dav{We also consider a {\em same-grid setting}, in which the input and output spectrograms are computed on the previous $M \times N$ support. More precisely, we compute $\mathbf{X}_1', \mathbf{X}_2'\in\mathbb{R}_+^{M\times N}$ with the same window sizes $W_1$ and $W_2$, respectively, but now each with hop size $2\,\mathrm{ms}$ and frequency spacing $2\,\mathrm{Hz}$.} For comparison, we compute the same-grid UOT barycenter of $\mathbf{X}'_1$ and $\mathbf{X}'_2$ as well as the same-grid NMF-fused spectrograms, with the same hyper-parameters as above. Additionally, we also compute the geometric-mean (GM) fused spectrogram \cite{cheung1991combined} given by
\begin{equation}
\mathbf{X}_G=\bigl(\mathbf{X}'_1\odot \mathbf{X}'_2\bigr)^{1/2}\in\mathbb{R}_+^{M\times N}.
\label{eq:geometric-mean}
\end{equation}
\dav{Finally, we also compare the results of the latter methods with time-frequency reassignment applied to $\mathbf{X}'_1$ and $\mathbf{X}'_2$ \cite{flandrin2018time}, yielding the reassigned spectrograms $\mathbf{R}_1$ and $\mathbf{R}_2$.}

\dav{For both the different-grid and same-grid settings, we also compare the UOT barycenters with and without the overlap constraints in the cost matrices. We do not include standard OT barycenters in the reported comparisons because their computation took approximately one hour per signal, with cost matrices containing more than 100 million finite entries. Moreover, standard OT barycenters consistently exhibited poor localization performance compared with the other methods.}

\subsubsection{\dav{{Nearest neighbor operators}}}
In the end, all the compared spectrograms are assessed on a common t-f grid with a temporal spacing of $2\,\mathrm{ms}$ and a frequency spacing of approximately $2\,\mathrm{Hz}$. \dav{To accommodate the continuously defined packet frequency \(f^\ast\), onset and offset times \(t^{\mathrm{on}}\) and \(t^{\mathrm{off}}\), we rely on the following discretization operators.} Let $f, t \in \mathbb{R}$ denote a frequency and a time instant, respectively, and let $\mathcal{S} = \mathcal{F} \times \mathcal{T}$ be a t-f support containing $M\times N$ points $(f_m,t_n)$. We define the frequency and temporal nearest neighbor operators as
\begin{equation}
P_{\mathcal{F}}(f)\in\underset{f_m \in \mathcal{F}}{\mathrm{argmin}}\,|f-f_m|, \quad P_{\mathcal{T}}(t)\in\underset{t_n \in \mathcal{T}}{\mathrm{argmin}}\,|t - t_n|.
\label{eq:closest-freq}
\end{equation}
\dav{If multiple solutions exist}, we select the smaller one.

\subsection{Single time-frequency packets} \label{sec:tf-localization-single}

In our first experimental setting we generate signals consisting of a single sinusoidal packet with random frequency $f^\ast$, onset time $t^\mathrm{on}$, and offset time $t^\mathrm{off}$, and evaluate them on the t-f grid $\mathcal{S}=\mathcal{F}\times\mathcal{T}$.  Our \dav{first} goal is to evaluate the following two aspects separately:
\begin{itemize}
    \item[(i)] Frequency localization: how well does a spectrogram concentrate energy in frequency, around the sinusoid frequency?
    \item[(ii)] Temporal localization: how well does it concentrate energy in time, between the onset and offset instants?
\end{itemize}

\dav{We also evaluate how the energy is distributed within a packet's time-frequency representation. Indeed, some methods achieve excellent localization by concentrating energy on very few t-f points, or exhibit irregular energy distributions, which are undesirable behaviors. As such, we introduce two additional metrics:
\begin{itemize}
    \item[(i)] \(\mathcal{L}^2\)-error: how close is the energy distribution to an idealized uniform density?
    \item[(ii)] Total-variation (TV) error: how smooth is the energy distribution?
\end{itemize}
}
\subsubsection{\dav{{Frequency and time localization properties}}}
 To assess frequency localization, we define the tolerance interval 
\begin{equation}
\mathcal I_{\Delta_f}(f^\ast) = [P_\mathcal{F}(f^\ast)-\Delta_f,\; P_\mathcal{F}(f^\ast)+\Delta_f],
\end{equation}
where $\Delta_f\in\mathbb{R}_+$ controls the tolerance along the frequency axis. We define the error of energy concentration in frequency as
\begin{equation}
E_f(\mathbf{Z}, \Delta_f) =\frac{\sum_{m,n} Z_{mn}\,\mathbbm{1}\!\left[f_m\notin \mathcal I_{\Delta_f}(f^\ast)\right]}{\sum_{m,n} Z_{mn}},
\label{eq:error-frequency}
\end{equation}
where $\mathbf{Z} \in \mathbb{R}_{+}^{M \times N}$ denotes a spectrogram and, for a set $\mathcal{A}$, we denote
\begin{equation}
    \mathbbm{1}[x \notin \mathcal{A}] =
    \begin{cases}
        1 \quad \text{if } x \notin \mathcal{A},\\
        0 \quad \text{otherwise.}
    \end{cases}
\end{equation} 
$E_f$ measures the proportion of the spectrogram's energy 
supported outside the region of tolerance in frequency.  
Similarly, to assess temporal localization, we define the tolerance interval
\begin{equation}
\mathcal I_{\Delta_t}(t^{\mathrm{on}},t^{\mathrm{off}}) = [P_\mathcal{T}(t^{\mathrm{on}})-\Delta_t,\; P_\mathcal{T}(t^{\mathrm{off}})+\Delta_t],
\end{equation}
where $\Delta_t\in\mathbb{R}_+$ controls the tolerance along the time axis. We define the error of energy concentration in time as
\begin{equation}
E_t(\mathbf{Z}, \Delta_t) = \frac{ \sum_{m,n} Z_{mn}\,\mathbbm{1}\!\left[t_n\notin \mathcal I_{\Delta_t}(t^{\mathrm{on}},t^{\mathrm{off}})\right] }{ \sum_{m,n} Z_{mn}},
\label{eq:error-time}
\end{equation}
which measures the proportion of the spectrogram's energy 
supported outside of the region of tolerance in time.  

\subsubsection{\dav{{Interior energy distribution}}}\label{sec:single-interior-energy}

\dav{We also evaluate how energy is distributed within the idealized t-f support, defined as}
\begin{equation}
    \dav{\mathcal I_0 = \{P_\mathcal F(f^\ast)\}\times [P_\mathcal T(t^{\mathrm{on}}),P_\mathcal T(t^{\mathrm{off}})].}
\label{eq:single-true-tf-supp}
\end{equation}
\dav{We denote $p_{\mathbf Z}$ the (normalized) temporal energy supported on $\mathcal I_0$ of a spectrogram $\mathbf Z$.}
\dav{We compare this density to the uniform density on $\mathcal I_0$, denoted $p_0$, through the $\mathcal{L}_2$-error:}
\begin{equation}
    \dav{E_{\mathcal{L}_2}(\mathbf Z)=\Vert p_\mathbf{Z} - p_0 \Vert_2.}
\label{eq:l2-error-single}
\end{equation}
\dav{This error measures deviations from a uniform energy distribution within the packet's support. Similarly, we quantify the smoothness of each profile with their total variation (TV)}
\begin{equation}
    \dav{E_{\mathrm{TV}}(\mathbf{Z}) = \sum_{n} \vert p_\mathbf{Z}(n + 1) - p_\mathbf{Z}(n) \vert.}
\label{eq:tv-error-single}
\end{equation}

\subsubsection{\dav{{Results}}}

\dav{Table~\ref{tab:single-packet} reports the evaluation metrics for selected values of $\Delta_f$ and $\Delta_t$, averaged over 100 experiments.} \dav{Regarding frequency localization, the high-temporal resolution spectrogram $\mathbf X'_2$ exhibits the more robust error decrease. The UOT barycenters closely match the frequency localization performance of the high-frequency resolution spectrogram $\mathbf{X}'_1$, with only small relative differences. This shows that our method successfully preserves its frequency localization properties. Similar results are obtained with rank-$1$ NMF fusion (either same- or different-grid), whereas the GM and the other different-grid NMF results perform worse. In contrast, reassignment and larger-rank same-grid NMF results improve upon \(\mathbf{X}'_1\), with the most pronounced improvement obtained by reassigning $\mathbf X'_2$. Similar trends are observed for temporal localization, where $\mathbf X'_1$ exhibits by far the largest errors. The UOT barycenters closely match the localization of \(\mathbf{X}'_2\), with only small relative differences.}

\dav{Although reassignment achieves better time-frequency localization, this comes at the expense of energy distribution within the true support, producing the least regular and least uniform density among all methods. In contrast, UOT barycenters maintain relatively smooth and uniform profiles.}

\dav{Regarding computational cost, the GM and reassigned spectrograms are obtained at very small cost, whereas the UOT and NMF methods are more involved (but only in the order of seconds). The different-grid UOT barycenter with overlap constraints requires only \(0.88\,\mathrm{s}\) on average, around \(26\) times faster than its same-grid counterpart. In both the same-grid and different-grid scenarios, overlap constraints substantially reduce runtime while preserving localization and energy distribution properties.}

\dav{Overall, these experiments demonstrate that the UOT barycenters retain both the frequency localization of \(\mathbf{X}'_1\) and the temporal localization of \(\mathbf{X}'_2\), while maintaining relatively smooth and uniform energy distributions. In particular, the different-grid UOT barycenter with overlap constraints offers the most favorable compromise between localization, energy distribution, and runtime among the fusion methods considered.}


\begin{table*}[t]
\centering
\caption{\dav{Single packet experiment. Comparison of temporal and frequency localization, as well as computational load. For the frequency (resp. temporal) localization error, we notice that the UOT barycenters closely follow that of \(\mathbf{X}'_1\) (resp. \(\mathbf{X}'_2\)).}}
\label{tab:single-packet}
\footnotesize
\setlength{\tabcolsep}{3pt}
\begin{tabular}{l*{15}{c}}
\toprule

& \multicolumn{2}{c}{Inputs}
& \multicolumn{4}{c}{UOT}
& \multicolumn{1}{c}{GM}
& \multicolumn{2}{c}{Reassignment}
& \multicolumn{6}{c}{NMF} \\

\cmidrule(lr){2-3}
\cmidrule(lr){4-7}
\cmidrule(lr){8-8}
\cmidrule(lr){9-10}
\cmidrule(lr){11-16}

&
&
& \multicolumn{2}{c}{Same grid}
& \multicolumn{2}{c}{Diff. grid}
&
&
&
& \multicolumn{3}{c}{Same grid}
& \multicolumn{3}{c}{Diff. grid} \\

\cmidrule(lr){4-5}
\cmidrule(lr){6-7}
\cmidrule(lr){11-13}
\cmidrule(lr){14-16}

& $\mathbf{X}'_1$ & $\mathbf{X}'_2$
& Overlap & No ov.
& Overlap & No ov.
& $\ve{X}_G$
& $\mathbf{R}_1$ & $\mathbf{R}_2$
& $K=1$ & $K=10$ & $K=100$
& $K=1$ & $K=10$ & $K=100$ \\

\midrule

\multicolumn{16}{l}{\textit{Frequency localization error (\%)}} \\

0 Hz
& 95.9 & 98.0
& 95.9 & 95.9 & 95.9 & 95.9
& 96.7
& 95.9 & 76.5
& 95.9 & 95.7 & 95.7
& 95.9 & 96.1 & 96.2 \\

5 Hz
& 71.8 & 85.8
& 72.3 & 72.3 & 71.9 & 71.9
& 77.4
& 68.4 & 42.5
& 71.8 & 70.6 & 70.6
& 71.8 & 73.8 & 74.0 \\

10 Hz
& 57.5 & 77.8
& 58.1 & 58.1 & 57.6 & 57.6
& 65.4
& 52.6 & 32.0
& 57.5 & 55.7 & 55.6
& 57.5 & 60.4 & 60.6 \\

15 Hz
& 39.7 & 66.4
& 40.5 & 40.5 & 39.9 & 39.9
& 49.4
& 34.3 & 22.1
& 39.7 & 37.4 & 37.4
& 39.7 & 43.4 & 43.5 \\

20 Hz
& 30.7 & 59.2
& 31.4 & 31.4 & 30.9 & 30.9
& 40.5
& 26.0 & 17.8
& 30.7 & 28.4 & 28.3
& 30.7 & 34.4 & 34.4 \\

35 Hz
& 12.5 & 34.8
& 13.1 & 13.1 & 12.7 & 12.7
& 19.1
& 11.3 & 8.4
& 12.5 & 10.8 & 10.8
& 12.5 & 14.5 & 14.3 \\

50 Hz
& 6.9 & 18.6
& 7.4 & 7.4 & 7.0 & 7.0
& 10.6
& 6.5 & 4.7
& 6.8 & 5.6 & 5.6
& 6.9 & 7.7 & 7.4 \\

\midrule

\multicolumn{16}{l}{\textit{Temporal localization error (\%)}} \\

0 ms
& 39.0 & 2.0
& 2.3 & 2.3 & 2.8 & 2.8
& 5.0
& 0.6 & 0.1
& 2.3 & 1.9 & 1.9
& 2.4 & 7.4 & 8.8 \\

2 ms
& 32.4 & 0.4
& 0.4 & 0.4 & 0.7 & 0.7
& 1.5
& 0.4 & 0.0
& 0.7 & 0.5 & 0.5
& 0.7 & 4.6 & 6.5 \\

4 ms
& 26.5 & 0.0
& 0.0 & 0.0 & 0.2 & 0.2
& 0.3
& 0.3 & 0.0
& 0.3 & 0.2 & 0.2
& 0.4 & 3.6 & 5.5 \\

6 ms
& 21.3 & 0.0
& 0.0 & 0.0 & 0.2 & 0.2
& 0.0
& 0.2 & 0.0
& 0.3 & 0.1 & 0.1
& 0.3 & 3.3 & 5.0 \\

8 ms
& 16.9 & 0.0
& 0.0 & 0.0 & 0.2 & 0.2
& 0.0
& 0.2 & 0.0
& 0.2 & 0.1 & 0.1
& 0.3 & 2.9 & 4.6 \\

\midrule

\multicolumn{16}{l}{\textit{Energy distribution within true t-f support}} \\

$L_2$ error
& 0.03 & 0.15
& 0.13 & 0.13 & 0.12 & 0.12
& 0.11
& 0.57 & 0.37
& 0.12 & 0.14 & 0.14
& 0.12 & 0.22 & 0.21 \\

TV error
& 0.05 & 0.23
& 0.20 & 0.20 & 0.19 & 0.19
& 0.16
& 1.06 & 0.64
& 0.17 & 0.20 & 0.20
& 0.17 & 0.43 & 0.38 \\

\midrule

\multicolumn{16}{l}{\textit{Computational performance}} \\

Runtime (s)
& 0.001 & 0.001
& 23.2 & 72.5 & 0.88 & 3.0
& 0.002
& 0.006 & 0.005
& 19.4 & 21.8 & 24.8
& 5.04 & 4.11 & 6.52 \\

Iterations
& --- & ---
& 228 & 228 & 134 & 133
& ---
& --- & ---
& 302 & 458 & 436
& 289 & 484 & 475 \\

\bottomrule
\end{tabular}
\end{table*}


\subsection{Mixtures of time-frequency packets} \label{sec:tf-localization-multi}

\subsubsection{\dav{{Simultaneous t-f localization}}}

We now consider signals composed of mixtures of sinusoidal packets of same amplitude. For a packet of frequency $f^\ast$, onset time $t^{\mathrm{on}}$, and offset time $t^{\mathrm{off}}$, we define its tolerance support as
\begin{equation}
\mathcal{I}_{\Delta_f,\Delta_t}(f^\ast,t^{\mathrm{on}},t^{\mathrm{off}})
=
\mathcal{I}_{\Delta_f}(f^\ast)\times \mathcal{I}_{\Delta_t}(t^{\mathrm{on}},t^{\mathrm{off}}).
\end{equation}
For a signal composed of $K$ packets with parameters $(\mathbf{f}^\ast, \mathbf{t}^\mathrm{on}, \mathbf{t}^\mathrm{off}) = \{(f_k^\ast,t_k^{\mathrm{on}},t_k^{\mathrm{off}})\}_{k=1}^{K}$, we define the mixture's tolerance support as
\begin{equation}
\mathcal{I}_{\Delta_f,\Delta_t}(\mathbf{f}, \mathbf{t}^\mathrm{on}, \mathbf{t}^\mathrm{off})
=
\bigcup_{k=1}^{K}
\mathcal{I}_{\Delta_f,\Delta_t}(f_k^\ast,t_k^{\mathrm{on}},t_k^{\mathrm{off}}).
\end{equation}
We then define the overall concentration error as
\begin{IEEEeqnarray}{l} \label{eq:overall}
    \hspace{-1em}E(\mathbf{Z}, \Delta_f, \Delta_t) = \nonumber\\
    \frac{
\sum_{m,n} Z_{mn}\,\mathbbm{1}\!\left[(f_m,t_n)\notin \mathcal{I}_{\Delta_f,\Delta_t}(\mathbf{f}^\ast, \mathbf{t}^\mathrm{on}, \mathbf{t}^\mathrm{off})\right]
}{
\sum_{m,n} Z_{mn}
},\nonumber\\
\end{IEEEeqnarray}
which quantifies the fraction of energy 
supported outside the union of enlarged ideal supports. 

\subsubsection{\dav{{Interior energy distribution}}}

\dav{For each packet $k$, we define its normalized temporal energy profile $p_{\mathbf{Z},k}$ as in Section \ref{sec:single-interior-energy}. We then average the $\mathcal{L}_2$ and TV errors over the $K$ packets, which characterize the energy distribution within the individual packet supports.}

\subsubsection{\dav{{Results}}}

\dav{Table~\ref{tab:multi_comparison} reports the overall time-frequency localization errors for selected frequency tolerances and $\Delta_t=0$, as well as profile errors and computational performance. The UOT barycenters outperform both input spectrograms separately. Reassignment achieves the lowest errors, particularly for $\mathbf{R}_2$, while large-rank same-grid NMF fusion exhibits performance comparable to UOT. In contrast, rank-$1$ NMF fusions generally perform worse than the inputs, and the geometric mean provides more limited improvements.}

\dav{As in the single-packet experiment, reassignment produces the least smooth and uniform energy distributions. The UOT barycenters maintain relatively smooth and uniform profiles. Rank-$1$ NMF fusion achieves comparable profile errors, whereas increasing the rank generally degrades these properties, especially in the different-grid setting.}

\dav{Regarding computational cost, the different-grid UOT barycenter with overlap constraints is the fastest fusion method requiring iterative optimization, with an average runtime of $0.93\,\mathrm{s}$. Overlap constraints substantially reduce runtime while preserving performance.}

\dav{Overall, these experiments show that UOT barycenters improve upon both inputs in time-frequency localization while maintaining regular and uniform energy distributions. The different-grid UOT barycenter with overlap constraints again offers a favorable compromise between localization, energy distribution, and computational cost.}


\begin{table*}[!t]
\centering
\caption{\dav{Multi packet experiment. Comparison of overall time-frequency localization errors, profile errors and computational performance. We notice that the UOT barycenters improve over the input spectrograms, with good energy distribution and low runtime.}}
\label{tab:multi_comparison}
\footnotesize
\setlength{\tabcolsep}{3pt}
\makebox[\textwidth][c]{%
\begin{tabular}{l*{15}{c}}
\toprule
& \multicolumn{2}{c}{Inputs}
& \multicolumn{4}{c}{UOT}
& \multicolumn{1}{c}{GM}
& \multicolumn{2}{c}{Reassignment}
& \multicolumn{6}{c}{NMF} \\

\cmidrule(lr){2-3}
\cmidrule(lr){4-7}
\cmidrule(lr){8-8}
\cmidrule(lr){9-10}
\cmidrule(lr){11-16}

&
&
& \multicolumn{2}{c}{Same grid}
& \multicolumn{2}{c}{Diff. grid}
&
&
&
& \multicolumn{3}{c}{Same grid}
& \multicolumn{3}{c}{Diff. grid} \\

\cmidrule(lr){4-5}
\cmidrule(lr){6-7}
\cmidrule(lr){11-13}
\cmidrule(lr){14-16}

& $\mathbf{X}'_1$ & $\mathbf{X}'_2$
& Overlap & No ov.
& Overlap & No ov.
& $\ve{X}_G$
& $\mathbf{R}_1$ & $\mathbf{R}_2$
& $K=1$ & $K=10$ & $K=100$
& $K=1$ & $K=10$ & $K=100$ \\

\midrule
\multicolumn{16}{l}{\textit{Overall localization error ($\Delta_t=0$) (\%)}} \\

0 Hz
& 96.8 & 97.7
& 95.8 & 95.8 & 95.8 & 95.8
& 96.6
& 95.2 & 80.8
& 97.8 & 95.6 & 95.5
& 97.8 & 96.0 & 96.2 \\

10 Hz
& 67.8 & 75.5
& 57.4 & 57.4 & 57.4 & 57.4
& 64.6
& 49.8 & 35.0
& 77.6 & 55.8 & 55.7
& 77.8 & 60.1 & 61.7 \\

20 Hz
& 49.7 & 56.1
& 32.3 & 32.3 & 32.4 & 32.4
& 40.8
& 24.9 & 18.2
& 63.1 & 31.1 & 31.0
& 63.3 & 35.8 & 37.4 \\

30 Hz
& 41.5 & 40.0
& 20.0 & 20.0 & 20.0 & 20.0
& 26.6
& 15.3 & 10.5
& 53.5 & 19.5 & 19.5
& 53.8 & 22.9 & 24.1 \\

40 Hz
& 38.2 & 27.5
& 14.1 & 14.1 & 14.4 & 14.4
& 19.3
& 11.4 & 6.5
& 46.4 & 14.1 & 14.1
& 46.7 & 16.5 & 17.4 \\

50 Hz
& 35.8 & 16.9
& 10.1 & 10.1 & 10.6 & 10.6
& 13.8
& 8.8 & 4.1
& 39.2 & 10.2 & 10.2
& 39.6 & 11.9 & 12.9 \\

\midrule

\multicolumn{16}{l}{\textit{Energy distribution within true t-f support}} \\

$L_2$ error
& 0.04 & 0.16
& 0.15 & 0.15 & 0.14 & 0.14
& 0.11
& 0.55 & 0.41
& 0.12 & 0.13 & 0.13
& 0.12 & 0.20 & 0.23 \\

TV error
& 0.05 & 0.26
& 0.23 & 0.23 & 0.22 & 0.22
& 0.17
& 1.05 & 0.82
& 0.19 & 0.21 & 0.21
& 0.19 & 0.37 & 0.45 \\

\midrule
\multicolumn{16}{l}{\textit{Computational performance}} \\

Runtime (s)
& 0.001 & 0.001
& 27.2 & 84.5 & 0.93 & 3.4
& 0.002
& 0.005 & 0.005
& 12.1 & 25.0 & 25.2
& 3.64 & 6.76 & 8.79 \\

Iterations
& --- & ---
& 198 & 198 & 123 & 134
& ---
& --- & ---
& 57 & 340 & 276
& 84 & 496 & 492 \\

\bottomrule
\end{tabular}%
}
\end{table*}

\begin{table*}[!t]
\centering
\caption{\dav{Speech experiment. Comparison of harmonic concentration errors and computational performance. We notice that the UOT barycenters closely follow the frequency localization of \(\mathbf{X}'_1\).}}
\label{tab:speech_comparison}
\footnotesize
\setlength{\tabcolsep}{3pt}
\makebox[\textwidth][c]{%
\begin{tabular}{l*{13}{c}}

\toprule
& \multicolumn{2}{c}{Inputs}
& \multicolumn{2}{c}{UOT}
& \multicolumn{1}{c}{GM}
& \multicolumn{2}{c}{Reassignment}
& \multicolumn{6}{c}{NMF} \\

\cmidrule(lr){2-3}
\cmidrule(lr){4-5}
\cmidrule(lr){6-6}
\cmidrule(lr){7-8}
\cmidrule(lr){9-14}

&
&
& 
& 
&
&
&
& \multicolumn{3}{c}{Same grid}
& \multicolumn{3}{c}{Diff. grid} \\

\cmidrule(lr){9-11}
\cmidrule(lr){12-14}

& $\mathbf{X}'_1$ & $\mathbf{X}'_2$
& Same grid & Diff. grid
& $\ve{X}_G$
& $\mathbf{R}_1$ & $\mathbf{R}_2$
& $K=1$ & $K=10$ & $K=100$
& $K=1$ & $K=10$ & $K=100$ \\

\midrule

\multicolumn{14}{l}{\textit{Harmonic concentration error (\%)}} \\

0 Hz
& 72.4 & 90.5
& 73.0 & 72.9
& 79.6
& 60.7 & 67.0
& 89.6 & 73.9 & 72.4
& 89.6 & 76.2 & 76.6 \\

8 Hz
& 34.6 & 72.0
& 35.6 & 35.5
& 46.9
& 25.7 & 33.3
& 71.3 & 37.8 & 35.3
& 71.3 & 41.7 & 41.6 \\

16 Hz
& 19.1 & 54.9
& 19.8 & 19.7
& 28.5
& 13.9 & 19.4
& 57.0 & 22.2 & 20.3
& 57.0 & 24.9 & 24.4 \\

23 Hz
& 11.9 & 39.6
& 12.3 & 12.2
& 18.3
& 8.4 & 12.3
& 45.1 & 14.2 & 12.8
& 45.1 & 16.0 & 15.3 \\

31 Hz
& 7.5 & 26.8
& 7.8 & 7.7
& 11.7
& 5.1 & 7.6
& 34.6 & 9.1 & 8.1
& 34.6 & 10.4 & 9.7 \\

39 Hz
& 4.7 & 16.4
& 4.9 & 4.8
& 7.1
& 3.0 & 4.2
& 25.4 & 5.6 & 4.9
& 25.3 & 6.5 & 6.0 \\

47 Hz
& 2.8 & 9.1
& 2.9 & 2.9
& 4.1
& 1.7 & 2.0
& 18.0 & 3.2 & 2.8
& 18.0 & 3.8 & 3.4 \\

55 Hz
& 1.7 & 4.8
& 1.7 & 1.7
& 2.4
& 1.0 & 1.1
& 12.6 & 1.9 & 1.6
& 12.6 & 2.3 & 2.0 \\

\midrule
\multicolumn{14}{l}{\textit{Computational performance}} \\

Runtime (s)
& 0.006 & 0.006
& 92.3 & 10.2
& 0.013
& 0.050 & 0.039
& 17.2 & 135 & 180
& 10.4 & 80.5 & 186 \\

Iterations
& --- & ---
& 245 & 142
& ---
& --- & ---
& 15 & 145 & 171
& 40 & 339 & 610 \\

\bottomrule
\end{tabular}%
}
\end{table*}


\section{Experiments with speech signals} \label{sec:real-signals}

\noindent We now apply our proposed fusion method to recorded speech signals. In Section \ref{sec:harmonic-concentration}, we introduce a metric based on the harmonic structure of voiced frames \cite{rabiner2007introduction} and use it to evaluate the frequency localization achieved by each spectrogram. We then consider temporal localization in Section \ref{sec:temporal-distribution}. In principle, a quantitative evaluation of temporal localization should measure how well each spectrogram isolates and distinguishes voiced frames, silence, transients, fricatives, and other speech events. However, such an analysis would require appropriate frame-level annotations, which are not available in our setting. We therefore limit the temporal evaluation to a qualitative visual analysis of the spectrograms.

\subsection{Speech dataset and implementation details}

\subsubsection{\dav{{Signal description}}}

We use the PTDB-TUG database \cite{pirker2011pitch} of speech samples for our evaluation. The database consists of recordings from both men and women along with pitch trajectories that will be used to define the harmonic supports. We randomly select $100$ speech signals with a variable duration of approximately $5\,\mathrm{s}$ using a $50/50$ split by gender. We append silence to the end of each signal so that all signals have the same duration. We downsample the signals to $f_s = 8\,\mathrm{kHz}$ to keep intelligible speech signals while maintaining a reduced runtime. For the target support, we select $5\,\mathrm{ms}$ of temporal spacing and approximately $8\,\mathrm{Hz}$ of frequency spacing. 

\subsubsection{\dav{{Experimental setting}}}

\dav{As in Section \ref{sec:imp-details}, we consider two grid scenarios for computation of the input spectrograms. In a different-grid setting, we compute $\mathbf{X}_1$ with a $100\,\mathrm{ms}$ window, $8\,\mathrm{Hz}$ frequency spacing and $75\,\%$ overlap. For $\mathbf{X}_2$, we use a $20\,\mathrm{ms}$ window, complete frequency sampling (resulting in $50\,\mathrm{Hz}$ spacing) and $5\,\mathrm{ms}$ of hop size. Like in Section \ref{sec:imp-details}, in the same-grid setting, the spectrograms $\mathbf{X}'_1$ and $\mathbf{X}'_2$ are over-sampled in time or frequency to achieve common frequency spacing and hop-size. We use the same experimental setup as in Section~\ref{sec:imp-details}, except that we omit the UOT barycenters without overlap constraints as we could not store the cost matrices in memory.}

\subsection{Harmonic concentration for voiced speech} \label{sec:harmonic-concentration}

Short segments of voiced speech can be approximated through a harmonic model \cite{rabiner2007introduction}. 
For a voiced segment, we can associate a fundamental frequency $f_0(t)$ at any given time $t$.
Let $\mathcal{S}= \mathcal{F} \times \mathcal{T}$ denote the target t-f support. At frame $t_n \in \mathcal{T}$, the ideal t-f representation of a speech signal localizes its energy over the frequency points
\begin{equation}
    \mathcal{H}(f_0(t_n)) = \bigcup_{k = 1}^{K_\mathrm{max}(t_n)} \{P_\mathcal{F}(kf_0(t_n))\} \subset \mathcal{F},
\end{equation}
where $P_\mathcal{F}(kf_0(t_n))$ defines the 
nearest neighbor in $\mathcal{F}$ to the $k$-th harmonic, see \eqref{eq:closest-freq}, and $K_\mathrm{max}(t_n)$ denotes the index of the last harmonic, usually set to the maximum integer $k$ verifying $kf_0(t_n) \leq f_s/2$. Since spectrograms spread energy in the frequency axis, we define an enlarged harmonic support of frequency points
\begin{IEEEeqnarray}{l}
    \mathcal{H}_{\Delta_f}(f_0(t_n)) =\nonumber\\
    \bigcup_{k = 1}^{K_\mathrm{max}(t_n)} \{f_m \in \mathcal{F}\,\vert\, 
    \vert f_m -P_\mathcal{F}(kf_0(t_n))\vert \leq \Delta_f \}
    \subset \mathcal{F},\nonumber\\
\end{IEEEeqnarray}
where $\Delta_f \geq 0$ controls the tolerance along the frequency axis. For a speech spectrogram $\mathbf{Z} \in \mathbb{R}_{+}^{M \times N}$, let $\mathcal{V}$ denote the set of time indices $n \in \{1, \ldots, N\}$ 
such that frame $t_n$ is voiced. With these notations, we introduce the error of harmonic energy concentration as
\begin{equation}
    E_H(\mathbf{Z}, \Delta_f) = \frac{\sum_{n\in \mathcal{V}} \sum_m Z_{mn} \mathbbm{1}[f_m \notin \mathcal{H}_{\Delta_f}(f_0(t_n))]}{\sum_{n \in \mathcal{V}}\sum_m Z_{mn}}.
\end{equation}
It measures the amount of energy that lies outside the enlarged harmonic support for voiced frames in a spectrogram. In the experiment, the set of voiced-frame indices  $\mathcal V$ is derived from the pitch annotations in PTDB-TUG, by retaining the frames 
with a valid estimate of $f_0(t_n)$.

\dav{Table~\ref{tab:speech_comparison} reports results for the harmonic energy concentration error as well as computational performance for the compared methods. As in the synthetic experiments, the UOT barycenters match the frequency localization of \(\mathbf{X}'_1\), while \(\mathbf{X}'_2\) exhibits substantially larger errors. Reassignment achieves the lowest errors, whereas the GM provides more limited improvements. Regarding NMF, increasing the rank generally improves this error, with better results achieved for same-grid NMF with rank \(K=100\). UOT exhibits lower runtimes when compared to NMF. Among these methods, different-grid UOT achieves the lowest runtime, with \(10\,\mathrm{s}\) per computation on average with nearly identical performance to its same-grid counterpart.}

\subsection{\dav{Qualitative analysis of speech spectrograms}}\label{sec:temporal-distribution}

\begin{figure}[!t]
\includegraphics[width=\columnwidth]{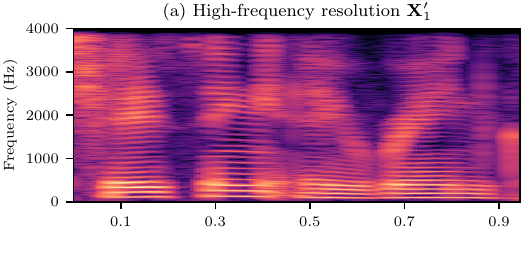}

\includegraphics[width=\columnwidth]{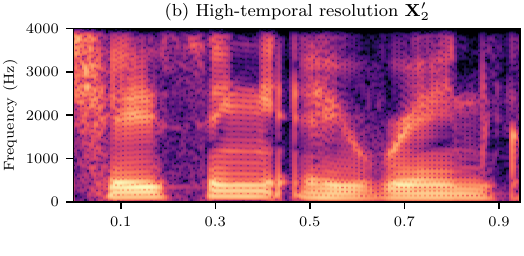}

\includegraphics[width=\columnwidth]{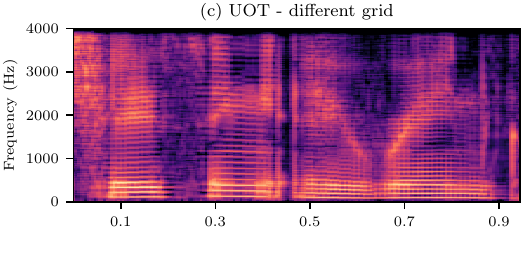}

\includegraphics[width=\columnwidth]{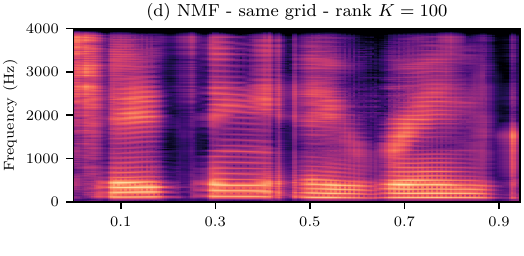}

\includegraphics[width=\columnwidth]{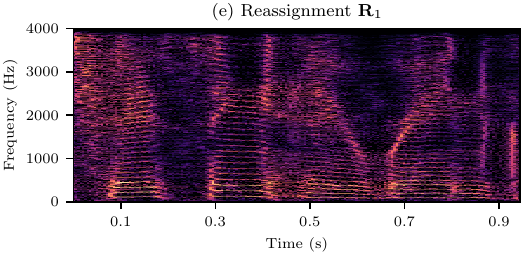}

\caption{Spectrograms computed from a male speech utterance saying ``Chasing a raincloud,'' from \cite{pirker2011pitch}. We display a $1$-second-long segment from the original signal. Magnitudes are displayed on a logarithmic scale.}
\label{fig:speech_comparison}
\end{figure}

\dav{We now qualitatively compare the two input spectrograms, the different-grid UOT barycenter, the rank-$100$ NMF same-grid result, and reassignment of $\mathbf{X}'_1$, as illustrated in Fig.~\ref{fig:speech_comparison}. These representations correspond to the same male speech utterance and allow us to examine properties not captured by the harmonic concentration metric.}

\dav{Regarding the temporal dynamics of speech, $\mathbf{X}'_1$ tends to smear energy over time, reducing the contrast between moments of low and high energy and blurring transitions between successive speech events. In comparison, $\mathbf{X}'_2$ captures sharper temporal boundaries and finer short-time variations, at the expense of frequency localization. The UOT barycenter appears to retain the harmonic structure of $\mathbf{X}'_1$ while recovering much of the temporal precision of $\mathbf X'_2$, particularly around abrupt transitions and narrow low-energy gaps. While this observation is only qualitative, it is consistent with the behavior observed on synthetic signals: the proposed fusion method preserves the temporal localization properties of the short-window spectrogram while maintaining the frequency concentration of the long-window one.}

\dav{Although rank-$100$ same-grid NMF achieves harmonic concentration errors comparable to UOT, its energy distribution appears more diffuse, particularly along the temporal axis, with less clear transitions. Conversely, reassignment achieves superior harmonic concentration but produces highly sparse and fragmented structures, potentially compromising the visual analysis of speech components.}

\dav{Overall, these observations suggest that UOT barycenters provide again a favorable compromise between frequency localization, temporal precision, energy distribution and runtime.}

\section{Conclusion}
\noindent We introduced an OT-based method for producing a super-resolution spectrogram based on the fusion of two or more spectrograms. By interpreting the input spectrograms as non-negative distributions, we combined their t-f resolutions by computing a UOT barycenter. We proposed new cost matrices that leverage t-f geometry and constrain energy transportation to neighboring t-f points. They yield well-localized super-resolution spectrograms and significantly reduce the computational runtime. Additionally, we designed a block MM algorithm to compute UOT barycenters without the need for entropic regularization. No such algorithm existed to compute UOT barycenters for input distributions with different supports. Experiments with synthetic and speech signals showed that the proposed method efficiently preserves the frequency localization of the high-frequency resolution spectrogram and the temporal localization of the high-temporal resolution spectrogram. \dav{Furthermore, it achieves a favorable balance between smooth energy distribution, computational efficiency, and localization accuracy compared with the other evaluated methods.} 

\section*{ACKNOWLEDGMENT}

We warmly thank Thomas Oberlin for stimulating discussions related to this work.

\printbibliography

@article{Bjoerkman2025,
  author       = {Bj{\"o}rkman, Anton and Sundstr{\"o}m, David and Jakobsson, Andreas and Elvander, Filip},
  title        = {Optimal Transport Regularization for Simulation-Informed Room Impulse Response Estimation},
  journal      = {IEEE Transactions on Signal Processing},
  volume       = {73},
  pages        = {5244--5256},
  year         = {2025},
}

@inproceedings{Fabiani2024,
  author       = {Fabiani, Linda and Schlecht, Sebastian J. and Elvander, Filip},
  title        = {Time-Frequency Audio Similarity Using Optimal Transport},
  booktitle    = {Proc. 58th Asilomar Conference on Signals, Systems, and Computers},
  pages        = {1414--1417},
  year         = {2024},
}

@inproceedings{Elvander2023,
  author       = {Elvander, Filip},
  title        = {Estimating Inharmonic Signals with Optimal Transport Priors},
  booktitle    = {Proc. IEEE International Conference on Acoustics, Speech, and Signal Processing (ICASSP)},
  pages        = {1--5},
  year         = {2023},
  organization = {IEEE},
}

@book{Flandrin1998,
  author       = {Flandrin, Patrick},
  title        = {Temps-Fr\'equence},
  edition      = {2},
  year         = {1998},
  publisher    = {Herm\`es},
}

@article{Cohen1989,
  author       = {Cohen, L.},
  title        = {Time-Frequency Distributions---A Review},
  journal      = {Proceedings of the IEEE},
  volume       = {77},
  number       = {7},
  pages        = {941--981},
  year         = {1989},
}

@inproceedings{Henderson2019,
  author       = {Trevor Henderson and Justin Solomon},
  title        = {Audio transport: a generalized portamento via optimal transport},
  booktitle    = {Proc. 22nd International Conference on Digital Audio Effects (DAFx-19)},
  year         = {2019},
}

@inproceedings{Flamary2016,
  author       = {Flamary, R{\'e}mi and F{\'e}votte, C{\'e}dric and Courty, Nicolas and Emiya, Valentin},
  title        = {Optimal spectral transportation with application to music transcription},
  booktitle    = {Advances in Neural Information Processing Systems (NIPS)},
  volume       = {29},
  pages        = {703--711},
  year         = {2016},
}

@book{oppenheim1999discrete,
  author       = {Oppenheim, Alan V. and Schafer, Ronald W. and Buck, John R.},
  title        = {Discrete-time signal processing},
  edition      = {2},
  year         = {1999},
  publisher    = {Prentice Hall},
}

@book{mallat1999wavelet,
  author       = {Mallat, St{\'e}phane},
  title        = {A wavelet tour of signal processing},
  edition      = {2},
  year         = {1999},
  publisher    = {Academic Press},
}

@inproceedings{cheung1991combined,
  author       = {Cheung, Shiufun and Lim, Jae S},
  title        = {Combined multi-resolution (wideband/narrowband) spectrogram},
  booktitle    = {Proc. IEEE International Conference on Acoustics, Speech, and Signal Processing (ICASSP)},
  pages        = {457--460},
  year         = {1991},
  organization = {IEEE},
}

@inproceedings{nam2010super,
  author       = {Nam, Juhan and Mysore, Gautham J and Ganseman, Joachim and Lee, Kyogu and Abel, Jonathan S},
  title        = {A super-resolution spectrogram using coupled {PLCA}},
  booktitle    = {Proc. Interspeech},
  pages        = {1696--1699},
  year         = {2010},
}

@article{kirbiz2014multiresolution,
  author       = {K{\i}rb{\i}z, S and G{\"u}nsel, Bilge},
  title        = {A multiresolution non-negative tensor factorization approach for single channel sound source separation},
  journal      = {Signal Processing},
  volume       = {105},
  pages        = {56--69},
  year         = {2014},
}

@book{boashash2015time,
  editor       = {Boashash, Boualem},
  title        = {Time-Frequency Signal Analysis: Methods and Applications},
  year         = {1992},
  publisher    = {Longman Cheshire},
}

@incollection{flandrin2018time,
  author       = {Flandrin, Patrick and Auger, Francois and Chassande-Mottin, Eric},
  editor       = {Papandreou-Suppappola, Antonia},
  title        = {Time-frequency reassignment: from principles to algorithms},
  booktitle    = {Applications in Time-Frequency Signal Processing},
  chapter      = {5},
  pages        = {179--203},
  year         = {2003},
  publisher    = {CRC Press},
  
}

@incollection{grossmann1990reading,
  author       = {Grossmann, Alexandre and Kronland-Martinet, Richard and Morlet, J},
  title        = {Reading and understanding continuous wavelet transforms},
  booktitle    = {Wavelets: Time-Frequency Methods and Phase Space},
  pages        = {2--20},
  year         = {1990},
  publisher    = {Springer},
}

@article{moca2021time,
  author       = {Moca, Vasile V. and B{\^a}rzan, Harald and Nagy-D{\u{a}}b{\^a}can, Adriana and Mure{\c{s}}an, Raul C.},
  title        = {Time-frequency super-resolution with superlets},
  journal      = {Nature Communications},
  volume       = {12},
  number       = {1},
  year         = {2021},
  note         = {Art. no. 337},
}

@article{cazelles2020wasserstein,
  author       = {Cazelles, Elsa and Robert, Arnaud and Tobar, Felipe},
  title        = {The {Wasserstein-Fourier} distance for stationary time series},
  journal      = {IEEE Transactions on Signal Processing},
  volume       = {69},
  pages        = {709--721},
  year         = {2021},
}

@inproceedings{valdivia2025audio,
  author       = {Valdivia, David and Renaud, Marien and Cazelles, Elsa and F{\'e}votte, C{\'e}dric},
  title        = {Audio signal interpolation using optimal transportation of spectrograms},
  booktitle    = {Proc. IEEE Statistical Signal Processing Workshop (SSP)},
  pages        = {1--5},
  year         = {2025},
  organization = {IEEE},
}

@article{griffin1984signal,
  author       = {Griffin, Daniel and Lim, Jae},
  title        = {Signal estimation from modified short-time {Fourier} transform},
  journal      = {IEEE Transactions on Acoustics, Speech, and Signal Processing},
  volume       = {32},
  number       = {2},
  pages        = {236--243},
  year         = {1984},
}

@inproceedings{elvander2017using,
  author       = {Elvander, Filip and Adalbj{\"o}rnsson, Stefan Ingi and Karlsson, Johan and Jakobsson, Andreas},
  title        = {Using optimal transport for estimating inharmonic pitch signals},
  booktitle    = {Proc. IEEE International Conference on Acoustics, Speech, and Signal Processing (ICASSP)},
  pages        = {331--335},
  year         = {2017},
  organization = {IEEE},
}

@inproceedings{fabiani2025joint,
  author       = {Fabiani, Linda and Schlecht, Sebastian J and Haasler, Isabel and Elvander, Filip},
  title        = {Joint spectrogram separation and {TDOA} estimation using optimal transport},
  booktitle    = {Proc. 33rd European Signal Processing Conference (EUSIPCO)},
  pages        = {221--225},
  year         = {2025},
  organization = {IEEE},
}

@article{peyre2020computationalot,
  author       = {Peyr{\'e}, Gabriel and Cuturi, Marco},
  title        = {Computational Optimal Transport: With Applications to Data Science},
  journal      = {Foundations and Trends in Machine Learning},
  volume       = {11},
  number       = {5--6},
  pages        = {355--607},
  year         = {2019},
}

@article{séjourné2023unbalancedoptimaltransporttheory,
  author       = {S{\'e}journ{\'e}, Thibault and Peyr{\'e}, Gabriel and Vialard, Fran{\c{c}}ois-Xavier},
  title        = {Unbalanced optimal transport, from theory to numerics},
  journal      = {Handbook of Numerical Analysis},
  volume       = {24},
  pages        = {407--471},
  year         = {2023},
}

@inproceedings{cuturi2014fast,
  author       = {Cuturi, Marco and Doucet, Arnaud},
  title        = {Fast computation of {W}asserstein barycenters},
  booktitle    = {Proc. 31st International Conference on Machine Learning (ICML)},
  series       = {Proceedings of Machine Learning Research},
  volume       = {32},
  pages        = {685--693},
  year         = {2014},
  publisher    = {PMLR},
}

@article{chizat2017scalingalgorithmsunbalancedtransport,
  author       = {Chizat, L{\'e}na{\"\i}c and Peyr{\'e}, Gabriel and Schmitzer, Bernhard and Vialard, Fran{\c{c}}ois-Xavier},
  title        = {Scaling algorithms for unbalanced optimal transport problems},
  journal      = {Mathematics of Computation},
  volume       = {87},
  number       = {314},
  pages        = {2563--2609},
  year         = {2018},
}

@article{POT,
  author       = {R{\'e}mi Flamary and Nicolas Courty and Alexandre Gramfort and Mokhtar Z. Alaya and Aur{\'e}lie Boisbunon and Stanislas Chambon and Laetitia Chapel and Adrien Corenflos and Kilian Fatras and Nemo Fournier and L{\'e}o Gautheron and Nathalie T.H. Gayraud and Hicham Janati and Alain Rakotomamonjy and Ievgen Redko and Antoine Rolet and Antony Schutz and Vivien Seguy and Danica J. Sutherland and Romain Tavenard and Alexander Tong and Titouan Vayer},
  title        = {{POT}: Python {O}ptimal {T}ransport},
  journal      = {Journal of Machine Learning Research},
  volume       = {22},
  number       = {78},
  pages        = {1--8},
  year         = {2021},
}

@article{razaviyayn2013unified,
  author       = {Razaviyayn, Meisam and Hong, Mingyi and Luo, Zhi-Quan},
  title        = {A unified convergence analysis of block successive minimization methods for nonsmooth optimization},
  journal      = {SIAM Journal on Optimization},
  volume       = {23},
  number       = {2},
  pages        = {1126--1153},
  year         = {2013},
}

@article{Leplat2022,
  author       = {Leplat, Valentin and Gillis, Nicolas and F{\'e}votte, C{\'e}dric},
  title        = {Multi-resolution beta-divergence {NMF} for blind spectral unmixing},
  journal      = {Signal Processing},
  volume       = {193},
  year         = {2022},
  note         = {Art. no. 108428},
}

@inproceedings{chapel2021unbalanced,
  author       = {Chapel, Laetitia and Flamary, R{\'e}mi and Wu, Haoran and F{\'e}votte, C{\'e}dric and Gasso, Gilles},
  title        = {Unbalanced optimal transport through non-negative penalized linear regression},
  booktitle    = {Advances in Neural Information Processing Systems (NeurIPS)},
  volume       = {34},
  pages        = {23270--23282},
  year         = {2021},
}

@inproceedings{umesh1999fitting,
  author       = {Umesh, Srinivasan and Leon Cohen and Douglas Nelson},
  title        = {Fitting the mel scale},
  booktitle    = {Proc. IEEE International Conference on Acoustics, Speech, and Signal Processing (ICASSP)},
  volume       = {1},
  pages        = {217--220},
  year         = {1999},
  organization = {IEEE},
}

@book{o1987speech,
  author       = {O'Shaughnessy, Douglas},
  title        = {Speech Communication: Human and Machine},
  year         = {1987},
  publisher    = {Addison-Wesley},
  address      = {Reading, MA},
}

@inproceedings{mcfee2015librosa,
  author       = {McFee, Brian and Raffel, Colin and Liang, Dawen and Ellis, Daniel P. W. and McVicar, Matt and Battenberg, Eric and Nieto, Oriol},
  title        = {{{librosa}: Audio and Music Signal Analysis in Python}},
  booktitle    = {Proc. 14th Python in Science Conference (SciPy)},
  pages        = {18--24},
  year         = {2015},
}

@inproceedings{pirker2011pitch,
  author       = {Pirker, Gregor and Wohlmayr, Michael and Petrik, Stefan and Pernkopf, Franz},
  title        = {A Pitch Tracking Corpus with Evaluation on Multipitch Tracking Scenario},
  booktitle    = {Proc. Interspeech},
  pages        = {1509--1512},
  year         = {2011},
}

@article{rabiner2007introduction,
  author       = {Rabiner, Lawrence R and Schafer, Ronald W},
  title        = {Introduction to digital speech processing},
  journal      = {Foundations and Trends in Signal Processing},
  volume       = {1},
  number       = {1-2},
  pages        = {1--194},
  year         = {2007},
}

@article{shafi2009techniques,
  author       = {Shafi, Imran and Ahmad, Jamil and Shah, Syed Ismail and Kashif, Faisal M},
  title        = {Techniques to obtain good resolution and concentrated time-frequency distributions: a review},
  journal      = {EURASIP Journal on Advances in Signal Processing},
  volume       = {2009},
  number       = {1},
  year         = {2009},
  note         = {Art. no. 673539},
}

\end{document}


\sloppy

\title{Supplementary material for ``Enhancing time-frequency resolution with optimal transport and barycentric fusion of multiple spectrograms"}

\author{David Valdivia, Elsa Cazelles, Cédric Févotte, \IEEEmembership{Fellow,~IEEE,}
\thanks{The authors are with IRIT, University of Toulouse, CNRS, France (email: firstname.lastname@irit.fr).}}
\maketitle

\section{Neighboring sets for t-f indices}\label{sec:neighboring}

\noindent In this section, we explicitly derive the index sets $\mathcal{O}_1(n_1)$ and $\mathcal{O}_2(m_2)$ of neighboring time frames of $t_{n_1}$ and frequency bins $f_{m_2}$, respectively. For spectrograms $\mathbf{X}_1 \in \mathbb{R}^{M_1 \times N_1}$ and $\mathbf{X}_2 \in \mathbb{R}^{M_2 \times N_2}$, we denote $W_1 \in \mathbb{N}$ and $W_2 \in \mathbb{N}$ their respective window sizes, in samples, and $H_1 \in \mathbb{N}$ and $H_2 \in \mathbb{N}$ their respective hop sizes, in samples. We denote $f_s$ the sampling rate, in Hz.

\subsection{Temporal overlap}
For a window of size $W \in \mathbb{N}$ centered at time $t \in \mathbb{R}_{+}$, we denote
\begin{equation}
    T(t, W) = \left[ t - \frac{W}{2f_s}, t + \frac{W}{2f_s} \right]
\end{equation}
its temporal support. As illustrated in Fig.~4 in the main paper, we consider that time frames $t_{n_1} \in \mathcal{T}_1$ and $t_{n} \in \mathcal{T} = \mathcal{T}_2$ do not capture common signal content whenever their temporal supports do not overlap. Overlap occurs when
\begin{equation}
\begin{aligned}
&\quad T(t_{n_1}, W_1) \cap T(t_{n}, W_2) \neq \emptyset \\
\Longleftrightarrow  &
\left\{
\begin{aligned}
t_{n_1} + \frac{W_1}{2f_s} \geq t_{n} - \frac{W_2}{2f_s} & \qquad\text{if } t_{n} \geq t_{n_1}, \nonumber \\
t_{n_1} - \frac{W_1}{2f_s} \leq t_{n} + \frac{W_2}{2f_s} & \qquad\text{if } t_{n} \leq t_{n_1}.  \\
\end{aligned}
\right.
\end{aligned}
\label{eq:temporal-overlap}
\end{equation}
Using (7), this condition becomes
\begin{equation}
\begin{aligned}
\Longleftrightarrow  &
\left\{
\begin{aligned}
\frac{(n_1-1)H_1}{f_s} + \frac{W_1}{2f_s} \geq \frac{(n-1)H_2}{f_s} - \frac{W_2}{2f_s} & \;\;\text{if } t_{n} \geq t_{n_1}  \\
\frac{(n_1-1)H_1}{f_s} - \frac{W_1}{2f_s} \leq \frac{(n-1)H_2}{f_s} + \frac{W_2}{2f_s} & \;\;\text{if } t_{n} \leq t_{n_1}  \\
\end{aligned}
\right. \\
\Longleftrightarrow & \; n \in \mathcal{O}_1(n_1),
\end{aligned}
\label{eq:derive-o-t}
\end{equation}
where
\begin{IEEEeqnarray}{ll}
\mathcal{O}_1(n_1) = \Bigl[& (n_1-1) \frac{H_{1}}{H_{2}} - \frac{1}{2H_2} (W_1 + W_2 ) + 1, \nonumber\\
& (n_1 - 1)\frac{H_{1}}{H_{2}}  + \frac{1}{2H_2} (W_1 + W_2 ) + 1\Bigr].
\end{IEEEeqnarray}

\subsection{Frequency overlap} \label{sec:overlap-freq}
\begin{figure}[t]
    \centering \includegraphics[width=\columnwidth]{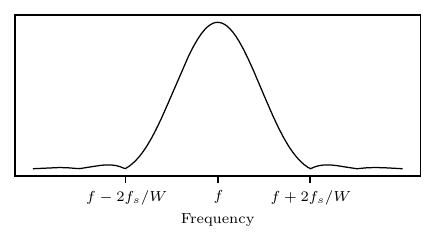}
    \caption{Fourier Transform of the Hann analysis window centered at $f$, in Hz. The main-lobe width is equal to $\frac{4f_s}{W}$, where $W$ is the length of the window.}
    \label{fig:freq-overlap}
\end{figure}

In Fig.~\ref{fig:freq-overlap} we display the power magnitude of the Fourier transform of the Hann window, centered at frequency $f \in \mathbb{R}$. We identify a main-lobe and multiple side-lobes of less intensity. While the side-lobes are present over the entire frequency axis, the energy of the main-lobe accounts for most of the window's energy. Thus, we define the \emph{effective} window's frequency support centered at $f \in \mathbb{R}$ as
\begin{equation}
F(f, W) = \Bigl[f - \frac{2f_s}{W}, f + \frac{2f_s}{W}\Bigr].
\label{eq:freq-support}
\end{equation}
Using a similar reasoning, frequency bins $f_{m_2} \in \mathcal{F}_2$ and $f_{m} \in \mathcal{F} = \mathcal{F}_1$ capture common signal content whenever their effective frequency supports overlap, i.e., if and only if
\begin{equation}
    F(f_{m_2}, W_2) \cap F(f_m, W_1) \neq \emptyset \Longleftrightarrow m \in \mathcal{O}_2(m_2),
\end{equation}
where the frequency neighboring set
\begin{IEEEeqnarray}{ll}
\mathcal{O}_2(m_2) = \Bigl[& (m_2 - 1) \frac{M_1 - 1}{M_2 - 1} - 4 (M_1 - 1)\Bigl(\frac{1}{W_1} + \frac{1}{W_2} \Bigr) + 1, \nonumber\\
& (m_2 - 1) \frac{M_1 - 1}{M_2 - 1} + 4 (M_1 - 1)\Bigl(\frac{1}{W_1} + \frac{1}{W_2} \Bigr) + 1\Bigr], \nonumber\\
\label{eq:freq-overlap-set}
\end{IEEEeqnarray}
is obtained with (8).

\section{\dav{Complexity analysis}}

\dav{In the general case, per-iteration complexity of Algorithm 1 is \(\mathcal{O}(IK + JK)\), as reported in Table~\ref{tab:complexity}. In the setting of the paper, \(I = M_1 N_1\), \(J = M_2 N_2\) and \(K = M_1 N_2\). When considering overlap constraints, the number of finite entries per row of each cost matrix is bounded above by \((W_1 + W_2) / H_2\) for \(\tilde{\mathbf{C}}_1\) and by \(8(M_1 - 1) (1 / W_1 + 1 / W_2)\) for \(\tilde{\mathbf{C}}_2\), leading to the reduced complexity reported in  Table~\ref{tab:complexity}.}

\dav{Moreover, letting $L$ denote the signal length in samples, with $L\approx N_iH_i$ for $i\in\{1,2\}$, and assuming standard STFT parameters $M_i\approx W_i/2$ and $H_i=W_i/2$, the complexity reduces to $\mathcal{O}(L(W_1+W_2)/W_2)$. In comparison, using unstructured cost matrices, the standard setting's per-iteration complexity is \(\mathcal{O}(L^2 W_1 / W_2)\). When varying the sampling rate \(f_s\), while keeping the signal, window, and hop durations fixed in seconds, the complexity scales as \(\mathcal{O}(f_s)\) with overlap costs and \(\mathcal{O}(f_s^2)\) for unstructured costs. Thus, under these respective assumptions, overlap constraints reduce the per-iteration complexity from quadratic to linear in both signal length and sampling rate.}

\begin{table}[t]
\centering
\caption{Computational complexity per iteration for different cost matrices.}
\label{tab:complexity}
\small
\renewcommand{\arraystretch}{1.15}
\setlength{\tabcolsep}{5pt}
\begin{tabular}{@{}ll@{}}
\toprule
Setting & Complexity \\
\midrule

General case
& $\mathcal{O}(IK+JK)$ \\

\midrule
\multicolumn{2}{@{}l}{\underline{General STFT setting}} \\
Unstructured
& $\mathcal{O}(M_1^2N_1N_2+M_1M_2N_2^2)$ \\
Overlap
& $\displaystyle\mathcal{O}\Bigl(
\frac{W_1+W_2}{H_2}M_1N_1$ \\
& $\displaystyle\qquad
+M_1\Bigl(\frac{1}{W_1}+\frac{1}{W_2}\Bigr)M_2N_2
\Bigr)$ \\

\midrule
\multicolumn{2}{@{}l}{\underline{Standard STFT parameters}} \\
\multicolumn{2}{@{}l}{\textit{Signal length $L$ scaling}} \\
Unstructured & $\mathcal{O}(L^2W_1/W_2)$ \\
Overlap & $\mathcal{O}(L(W_1+W_2)/W_2)$ \\

\addlinespace[4pt]
\multicolumn{2}{@{}l}{\textit{Sampling-rate $f_s$ scaling}} \\
Unstructured & $\mathcal{O}(f_s^2)$ \\
Overlap & $\mathcal{O}(f_s)$ \\
\bottomrule
\end{tabular}
\end{table}

\section{\dav{Sensitivity analysis of \(\eta\)}}

\dav{The marginal constraint relaxation parameter \(\eta\) in (13), in (30) and in (31), controls how much the UOT transport plans and barycenters preserve the original input weights. Table~\ref{tab:eta_sensitivity} reports all of the metrics used in the experiments. In the single-packet setting, while the frequency behavior is largely unaffected, the temporal localization and energy distribution deteriorate substantially with decreasing \(\eta\). Runtime on the other hand, is higher at \(\eta=1\). In the multi-packet setting, the trend in runtime and energy distribution is similar. However, concentration error is minimized at \(\eta=1\), rather than decreasing as \(\eta\) increases. For speech, the lowest harmonic concentration error is attained at \(\eta=1\). These results support the choice \(\eta=1\) for our experiments in the main paper.}

\begin{figure}[!t]
\includegraphics[width=\columnwidth]{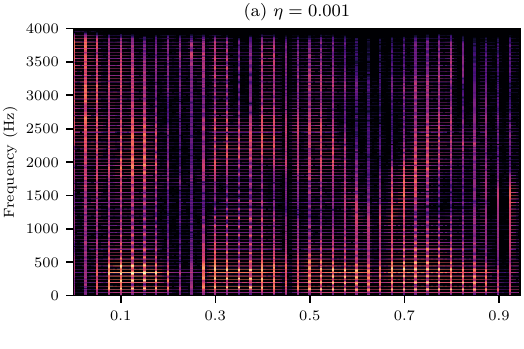}
\vspace{-.3cm}
\includegraphics[width=\columnwidth]{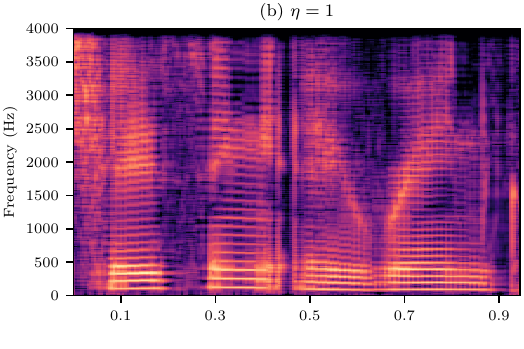}
\vspace{-.3cm}
\includegraphics[width=\columnwidth]{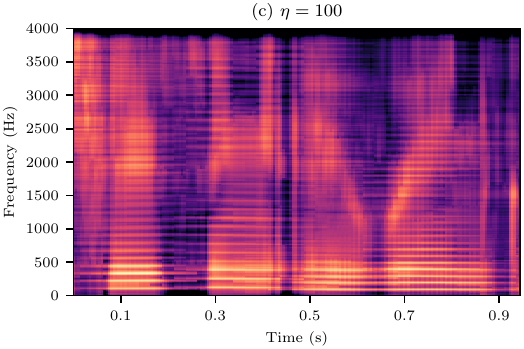}

\caption{Spectrograms computed from the speech excerpt of Fig.~5 in the main paper with varying \(\eta\).}
\label{fig:speech_eta}
\end{figure}

\dav{The influence of \(\eta\) is better observed in Fig.~\ref{fig:speech_eta}, which displays several speech spectrograms with different values of \(\eta\). We observe that very small values of \(\eta\)  produce sparse barycenters, consistent with the poor energy distribution metrics observed in the synthetic setting. On the other hand, higher values of \(\eta\) tend to produce more diffuse barycenters, explaining the improved energy distribution performance.}

\begin{table}[t]
\centering
\caption{Sensitivity of the different-grid UOT barycenter to the regularization parameter $\eta$ across single-packet, multi-packet, and speech experiments.}
\label{tab:eta_sensitivity}
\footnotesize
\setlength{\tabcolsep}{4pt}
\makebox[\columnwidth][c]{%
\begin{tabular}{l*{8}{r}}
\toprule
& \multicolumn{8}{c}{$\eta$} \\
\cmidrule(lr){2-9}
& $10^{-5}$ & $10^{-4}$ & $0.001$ & $0.01$
& $0.1$ & $1$ & $10$ & $100$ \\
\midrule

\multicolumn{9}{l}{\textbf{Single-packet experiments}} \\
\addlinespace[2pt]
\multicolumn{9}{l}{\textit{Frequency localization error (\%)}} \\
\cmidrule(lr){1-9}
$0$ Hz
& 96.1 & 96.2 & 96.1 & 96.1 & 96.0 & 95.9 & 95.9 & 95.9 \\
$5$ Hz
& 73.4 & 74.2 & 73.6 & 73.7 & 72.8 & 71.9 & 71.8 & 71.8 \\
$10$ Hz
& 59.7 & 61.0 & 60.1 & 60.1 & 58.8 & 57.6 & 57.4 & 57.5 \\
$15$ Hz
& 42.4 & 44.0 & 42.8 & 42.9 & 41.4 & 39.9 & 39.7 & 39.7 \\
$20$ Hz
& 33.4 & 34.9 & 33.6 & 33.8 & 32.4 & 30.9 & 30.7 & 30.7 \\
$35$ Hz
& 14.4 & 15.8 & 14.7 & 14.8 & 13.9 & 12.7 & 12.5 & 12.5 \\
$50$ Hz
& 7.9 & 8.7 & 8.2 & 8.3 & 7.9 & 7.0 & 6.8 & 6.8 \\

\addlinespace[3pt]
\multicolumn{9}{l}{\textit{Temporal localization error (\%)}} \\
\cmidrule(lr){1-9}
$0$ ms
& 43.7 & 38.9 & 34.6 & 31.3 & 14.9 & 2.8 & 2.2 & 2.3 \\
$2$ ms
& 38.7 & 34.0 & 28.4 & 25.3 & 12.0 & 0.7 & 0.4 & 0.5 \\
$4$ ms
& 30.0 & 26.2 & 21.7 & 19.7 & 10.9 & 0.3 & 0.1 & 0.1 \\
$6$ ms
& 24.8 & 21.6 & 17.8 & 16.2 & 10.1 & 0.2 & 0.0 & 0.1 \\
$8$ ms
& 21.3 & 18.4 & 15.0 & 13.7 & 9.2 & 0.2 & 0.0 & 0.1 \\

\addlinespace[3pt]
\multicolumn{9}{l}{\textit{Energy distribution within true t-f support}} \\
\cmidrule(lr){1-9}
$L_2$ error
& 0.79 & 0.78 & 0.72 & 0.63 & 0.18 & 0.12 & 0.11 & 0.11 \\
TV error
& 1.33 & 1.33 & 1.29 & 1.14 & 0.33 & 0.19 & 0.17 & 0.17 \\

\addlinespace[3pt]
\multicolumn{9}{l}{\textit{Computational performance}} \\
\cmidrule(lr){1-9}
Runtime (s)
& 0.06 & 0.10 & 0.19 & 0.53 & 0.48 & 0.88 & 0.44 & 0.38 \\
Iterations
& 9 & 15 & 29 & 81 & 75 & 134 & 61 & 58 \\

\midrule
\multicolumn{9}{l}{\textbf{Multi-packet experiments}} \\
\addlinespace[2pt]
\multicolumn{9}{l}{\textit{Overall localization error (\%)}} \\
\cmidrule(lr){1-9}
$0$ Hz
& 97.0 & 97.0 & 96.8 & 96.7 & 96.2 & 95.8 & 96.0 & 96.5 \\
$10$ Hz
& 69.1 & 69.2 & 67.7 & 66.8 & 61.5 & 57.4 & 59.2 & 64.7 \\
$20$ Hz
& 50.9 & 50.8 & 48.8 & 47.2 & 38.8 & 32.4 & 35.2 & 43.3 \\
$30$ Hz
& 42.4 & 41.7 & 39.6 & 37.6 & 27.5 & 20.1 & 23.1 & 31.7 \\
$40$ Hz
& 38.2 & 37.0 & 35.1 & 33.0 & 22.4 & 14.4 & 17.2 & 25.2 \\
$50$ Hz
& 35.3 & 33.6 & 32.0 & 29.7 & 18.8 & 10.6 & 13.1 & 20.0 \\

\addlinespace[3pt]
\multicolumn{9}{l}{\textit{Energy distribution within true t-f support}} \\
\cmidrule(lr){1-9}
$L_2$ error
& 0.77 & 0.76 & 0.71 & 0.62 & 0.19 & 0.14 & 0.12 & 0.12 \\
TV error
& 1.28 & 1.27 & 1.24 & 1.09 & 0.34 & 0.22 & 0.20 & 0.19 \\

\addlinespace[3pt]
\multicolumn{9}{l}{\textit{Computational performance}} \\
\cmidrule(lr){1-9}
Runtime (s)
& 0.07 & 0.11 & 0.21 & 0.58 & 0.51 & 0.94 & 0.84 & 0.41 \\
Iterations
& 10 & 14 & 29 & 80 & 70 & 130 & 123 & 56 \\
\midrule
\multicolumn{9}{l}{\textbf{Speech experiments}} \\
\addlinespace[2pt]
\multicolumn{9}{l}{\textit{Harmonic concentration error (\%)}} \\
\cmidrule(lr){1-9}
$0$ Hz
& 73.1 & 73.7 & 74.8 & 74.8 & 73.3 & 72.9 & 74.1 & 78.1 \\
$8$ Hz
& 36.1 & 37.4 & 39.7 & 39.4 & 36.1 & 35.5 & 37.7 & 45.6 \\
$16$ Hz
& 20.6 & 21.8 & 23.9 & 23.6 & 20.2 & 19.7 & 21.6 & 28.7 \\
$23$ Hz
& 12.9 & 13.8 & 15.4 & 15.3 & 12.6 & 12.2 & 13.7 & 19.3 \\
$31$ Hz
& 8.3 & 8.9 & 10.1 & 10.0 & 8.1 & 7.7 & 8.9 & 13.1 \\
$39$ Hz
& 5.2 & 5.5 & 6.2 & 6.2 & 5.1 & 4.8 & 5.6 & 8.6 \\
$47$ Hz
& 3.1 & 3.2 & 3.6 & 3.6 & 3.0 & 2.9 & 3.4 & 5.5 \\
$55$ Hz
& 1.8 & 1.9 & 2.1 & 2.1 & 1.8 & 1.7 & 2.0 & 3.5 \\

\addlinespace[3pt]
\multicolumn{9}{l}{\textit{Computational performance}} \\
\cmidrule(lr){1-9}
Runtime (s)
& 0.78 & 0.77 & 1.24 & 3.70 & 4.49 & 10.2 & 15.5 & 3.49 \\
Iterations
& 10 & 10 & 16 & 50 & 60 & 142 & 207 & 46 \\

\bottomrule
\end{tabular}%
}
\end{table}

\section{Target support customization} \label{sec:mel}

Previously, we considered the target support $\mathcal S=\mathcal F_1\times\mathcal T_2$ with the cost matrices in (27) and (28). To provide alternative target supports, appropriate cost matrices must be defined. In this section, we present an illustrative example that can serve as a template for other support configurations.
\subsection{Mel-frequency cost matrices}
\begin{figure}
\centering
\subfloat[]{
  \includegraphics[width=\columnwidth]{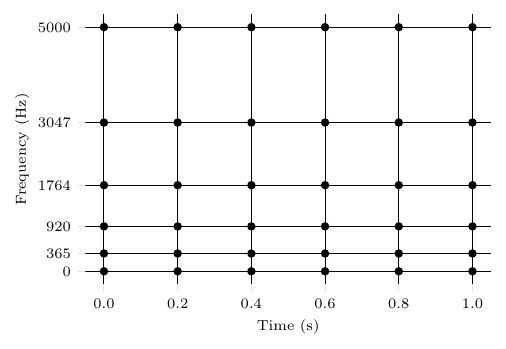}
}\\[-8mm]
\caption{Time-frequency support with frequency sampling defined on a mel scale, with 6 mel-spaced bins.}
\label{fig:mel-support}
\end{figure}

We replace the frequency sampling of the target support with a mel-frequency scale, as commonly done in speech applications.
We consider O'Shaughnessy's formula \cite{o1987speech} to convert hertz $f$ to mel $m$ as
\begin{equation}
    m = \mathrm{mel}(f) = 2595 \log_{10}\Bigl(1 + \frac{f}{700}\Bigr),
\label{eq:mel}
\end{equation}
and its inverse formula
\begin{equation}
    \mathrm{mel}^{-1}(m) = 700 \Bigl(10^{m/2595} - 1\Bigr).
\label{eq:imel}
\end{equation}
The mel-frequency bins are built as follows. Let $M \in \mathbb{N}$ be the number of mel bands, selected by the user. Let $m_r = \mathrm{mel}(f_s/2)$. Consider $M$ uniformly spaced values from $0$ to $m_r$, in the mel domain. Then, map each mel into the frequency domain using the inversion formula \eqref{eq:imel}. This results in the set 
\begin{equation}
    \mathcal{M} = \Bigl\{f_m:= \mathrm{mel}^{-1}\Bigl(\frac{m-1}{M-1} m_r\Bigr)\Bigr\}_{m=1}^{M}
\label{eq:mel-support}
\end{equation}
of $M$ frequencies regularly spaced in mel and expressed in Hertz.
Fig.~\ref{fig:mel-support} illustrates a support with frequency samplings distributed according to this construction.

Let $\mathbf X_1\in\mathbb R_{+}^{M_1\times N_1}$ and $\mathbf X_2\in\mathbb R_{+}^{M_2\times N_2}$ be two spectrograms with t-f supports $\mathcal S_1 = \mathcal{F}_1 \times \mathcal{T}_1$ and $\mathcal S_2 = \mathcal{F}_2 \times \mathcal{T}_2$, respectively. We consider the target support
\begin{equation}
\mathcal S=\mathcal M\times\mathcal T_2,
\label{eq:target-mel}
\end{equation}
where the $M$ frequency samplings are defined on a mel scale and the time samplings are taken from $\mathcal{S}_2$.

In the main paper, we considered the target support $\mathcal{S}=\mathcal{F}_1\times\mathcal{T}_2$. Accordingly, overlap sets only had to be defined between time frames $t_{n_1}\in\mathcal{T}_1$ and $t_n\in\mathcal{T}=\mathcal{T}_2$, and between frequency bins $f_{m_2}\in\mathcal{F}_2$ and $f_m\in\mathcal{F}=\mathcal{F}_1$. After replacing the target frequency sampling, new overlap sets must be introduced between $f_{m_1}\in\mathcal{F}_1$ and $f_m\in\mathcal{M}$, and between $f_{m_2}\in\mathcal{F}_2$ and $f_m\in\mathcal{M}$.

Assuming that the super-resolution spectrogram inherits the frequency resolution of $\mathbf{X}_1$, we define, for each frequency $f_m\in\mathcal{M}$, its effective frequency support as $F(f_m,W_1)$, see \eqref{eq:freq-support}. Using a reasoning similar to that in Section \ref{sec:overlap-freq}, overlap between $f_{m_2}\in\mathcal{F}_2$ and $f_m\in\mathcal{M}$ occurs if and only if
\begin{equation}
    F(f_{m_2}, W_2) \cap F(f_m, W_1) \neq \emptyset \Longleftrightarrow m \in \mathcal{O}'_2(m_2),
\end{equation}
where the neighboring-frequency set $\mathcal{O}'_2(m_2)$ is obtained by applying the same construction as in Section~\ref{sec:neighboring}. A similar set $\mathcal{O}'_1(m_1)$ is obtained by replacing $W_2$ with $W_1$ in the construction. With the notations above, we introduce the overlap cost matrix $\mathbf{C}'_1 \in \mathbb{R}_{+}^{M_1N_1 \times MN_2}$ between $\mathcal{S}_1$ and $\mathcal{S}$ defined entrywise as
\begin{equation}
C'_{1,i_1i}=
\begin{cases}
c_1(s_{i_1}, s_{i}), & \text{if } n\in\mathcal{O}_1(n_1) \text{ and } m \in \mathcal{O}'_1(m_1),\\
+\infty, & \text{otherwise.}
\end{cases}
\label{eq:cost-overlap-time-mel}
\end{equation}
Similarly, we introduce the overlap cost matrix $\mathbf{C}'_2 \in \mathbb{R}_{+}^{M_2N_2 \times MN_2}$ between $\mathcal{S}_2$ and $\mathcal{S}$ defined entrywise as
\begin{equation}
C'_{2,i_2i}=
\begin{cases}
c_2(s_{i_2}, s_{i}), & \text{if } n=n_2 \text{ and } m\in\mathcal{O}'_2(m_2),\\
+\infty, & \text{otherwise.}
\end{cases}
\label{eq:cost-overlap-freq-mel}
\end{equation}
\subsection{Application to a speech sample}
\begin{figure}[t]
\centering
\subfloat[]{
  \includegraphics[width=\columnwidth]{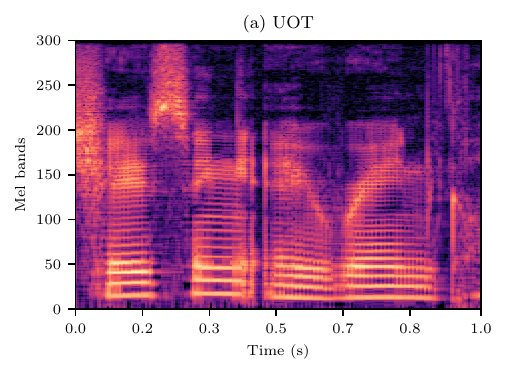}
  \label{fig:speech-mel-super-resolution}
}\\[-8mm]
\subfloat[]{
  \includegraphics[width=\columnwidth]{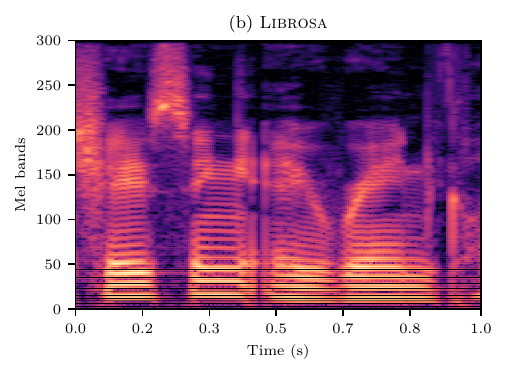}
  \label{fig:speech-mel-comparison}
}\\[-8mm]
\caption{Mel spectrograms computed from the speech excerpt of Fig.~5 in the main paper, with $M=300$ mel bands. Amplitude is displayed on a logarithmic scale. (a) UOT super-resolution mel spectrogram. (b) Reference mel spectrogram obtained using \textsc{Librosa} \cite{mcfee2015librosa}. The black horizontal lines indicate empty mel bands receiving no energy.}
\label{fig:speech-mel-results}
\end{figure}

We consider cost matrices defined above for the speech signal shown in Fig.~5 of the main document. We downsample the original signal to $f_s=22050\,\mathrm{Hz}$. We compute a high-frequency resolution spectrogram $\mathbf{X}_1$ with window size $W_1 = 100\,\mathrm{ms}$, frequency spacing of $8\,\mathrm{Hz}$ and $75\,\%$ overlap, which corresponds to $25\,\mathrm{ms}$ of hop size. We compute a high-temporal resolution spectrogram $\mathbf{X}_2$ with window size $W_2 = 20\,\mathrm{ms}$, temporal spacing of $5\,\mathrm{ms}$ and a complete frequency sampling, which corresponds to $50\,\mathrm{Hz}$. For the target support, we consider the setting $\mathcal{S} =\mathcal{M} \times \mathcal{T}_2$ from \eqref{eq:target-mel}. We select $M=300$ mel bands. We compute a UOT barycenter $\mathbf{X}$ between $\mathbf{X}_1$ and $\mathbf{X}_2$ using the cost matrices \eqref{eq:cost-overlap-time-mel} and \eqref{eq:cost-overlap-freq-mel} and the same experimental parameters as in Sections V and VI. The result is displayed in Fig.~\ref{fig:speech-mel-results}(a). 
We observe that the resulting super-resolution mel spectrogram provides a good t-f localization directly on the mel frequency axis.

For comparison, we compute a third spectrogram $\mathbf{X}_3$ with window size $W_3 = 40\,\mathrm{ms}$ providing intermediate t-f resolution. Its hop size and frequency spacing are set to $5\,\mathrm{ms}$ and $8\,\mathrm{Hz}$, respectively, and thus uses the frequency sampling of $\mathbf{X}_1$ and the temporal sampling of $\mathbf{X}_2$. We project this spectrogram on the mel frequency axis using a commonly used pipeline available in Librosa \cite{mcfee2015librosa}. The result is shown in Fig.~\ref{fig:speech-mel-results}(b). First, we observe that the mel spectrogram contains empty mel bands, a phenomenon arising when the number of mel bands $M$ is too large in comparison with the number of frequency bins. This example shows the potential of our method to compute well-defined mel spectrograms of arbitrary size. Second, we notice that the energy in $\mathbf{X}_3$ is spread in time and frequency when compared to UOT. This suggests that super-resolution is preserved even for arbitrary t-f supports.

\subsection{Summary}
In general, to adapt the construction of a UOT super-resolution spectrogram, we may follow these steps:
\begin{enumerate}
    \item Define an arbitrary target support $\mathcal{S}=\mathcal{F} \times \mathcal{T}$.
    \item Define the overlap sets between the pairs $(\mathcal{F}, \mathcal{F}_1)$, $(\mathcal{F}, \mathcal{F}_2)$, $(\mathcal{T}, \mathcal{T}_1)$ and $(\mathcal{T}, \mathcal{T}_2)$.
    \item Use these sets to construct cost matrices between the support pairs $(\mathcal{S}, \mathcal{S}_1)$ and $(\mathcal{S}, \mathcal{S}_2)$.
    \item Apply Algorithm 1 of the main paper using these cost matrices.
\end{enumerate}
Notably, this may be used to compute barycenters of more than two spectrograms, as discussed in Section IV-D of the main paper.

\printbibliography